\documentclass[11pt]{article}
\usepackage{amssymb,latexsym, amsmath}
\usepackage{amscd}

\makeatletter \@addtoreset{figure}{section}
\def\thefigure{\thesection.\@arabic\c@figure}
\def\fps@figure{h, t}
\@addtoreset{table}{bsection}
\def\thetable{\thesection.\@arabic\c@table}
\def\fps@table{h, t}
\@addtoreset{equation}{section}

\newif\ifamsfonts
\amsfontstrue \ifamsfonts
\font\twlbbb=msbm10 scaled\magstep1 \font\egtbbb=msbm8
\font\sixbbb=msbm6
\newfam\mathbbfam
\textfont\mathbbfam=\twlbbb \scriptfont\mathbbfam=\egtbbb
\scriptscriptfont\mathbbfam=\sixbbb
\makeatother

\newtheorem{example}{Example}[section]
\newtheorem{remark}{Remark}[section]
\newtheorem{theorem}{Theorem}[section]
\newtheorem{corollary}{Corollary}[section]
\newtheorem{definition}{Definition}[section]
\newtheorem{conjecture}{Conjecture}[section]

\newtheorem{question}{Question}[section]

\newcommand{\reg}{\mathrm{reg\;}}
\newcommand{\rank}{\mathrm{rank\;}}
\newcommand{\ddim}{\mathrm{ddim\;}}
\newcommand{\dind}{\mathrm{dind\;}}
\newcommand{\corank}{\mathrm{corank\;}}

\newcommand{\ann}{\mathrm{ann}}
\newcommand{\ad}{\mathrm{ad}}
\newcommand{\Ad}{\mathrm{Ad}}
\newcommand{\Span}{\mathrm{span\,}}
\newcommand{\diag}{\mathrm{diag}}
\newcommand{\dist}{\mathrm{dist}}
\newcommand{\rk}{\mathrm{rank\;}}
\newcommand{\ind}{\mathrm{ind\;}}
\newcommand{\const}{\mathrm{const}}

\begin{document}

\title{Integrable geodesic flows on Riemannian manifolds: Construction and Obstructions
}

\author{Alexey V. Bolsinov \footnote{e-mail:  bolsinov$@$mech.math.msu.su},
Bo\v zidar Jovanovi\' c  \footnote{e-mail: bozaj$@$mi.sanu.ac.yu} \\
\\
Department of Mechanics and Mathematics \\
Moscow State University\\
119992, Moscow, Russia \\
and\\
Mathematical Institute SANU\\
Kneza Mihaila 35, 11000 Belgrade, Serbia}

\date{}
\maketitle

\abstract
This paper is a review of recent and classical results on
integrable geodesic flows on Riemannian manifolds and topological
obstructions to integrability. We also discuss some open problems.
\endabstract

\tableofcontents

\section{Introduction}

\subsection{Geodesics on Riemannian manifolds}

We start with basic definitions and the statement of the problem to which our paper is devoted.

The first notion is {\it a geodesic line} on a manifold. To introduce it, consider the following problem.
Imagine a point moving on a two-dimensional surface in three-dimensional space. What is its trajectory,
if there is no external force acting on the point?  The
classical mechanics gives the following answer: the point will move with
velocity of constant absolute value and in such a way that the
acceleration (as a vector in $\Bbb R^3$) is always orthogonal to the
surface. The trajectory of the point is called {\it a geodesic line}.
It is not hard to verify that such a motion is described by a system
of two second-order differential equations. A less obvious fact is
that the geodesics are not changed under transformations which do not
touch the interior geometry of the
surface, or, in the language of differential geometry, the induced Riemannian metric.

The other approach to define geodesics uses the interior geometry
from the very beginning. Consider the following problem.
Given two points on a surface, find the shortest curve on the
surface connecting them.
Such a curve  (if it exists) is just a geodesic.
In general, there may exist several such curves or no one. However, if the
points are sufficiently close, then the desired curve always exists and
is unique. Thus the geodesics can be characterized as locally shortest
lines on the surface. This interpretation allows one to define geodesics
on an arbitrary manifold endowed with a Riemannian metric.

However, the first approach can be applied for the general case as
well.  Geodesics are the trajectories of points moving by inertia. In
other  words, the acceleration of a point must be identically zero.
One should only explain that in this case by the acceleration  we mean
the {\it covariant} derivative of the velocity vector. Let us recall
that such a derivative is naturally defined on every Riemannian manifold
and it is called {\it Levi-Civita connection}.

It can be easily shown that the geodesics are described by a system of
second-order differential equations which can be written in the
Hamiltonian form.
Let $M$ be a smooth manifold with a Riemannian metric $g=(g_{ij})$.
Consider an arbitrary local coordinate system $x^1,\dots,x^n$ and pass
from velocities $\dot x^i$ to momenta
$p_j$ by using the standard transformation $p_j= g_{ij}\dot x^i$.
Then in the new coordinates $x^i, p_i$
$(i=1,\dots,n)$ the equations of geodesics read:
\begin{equation}
\frac{dx^i}{dt}=\frac{\partial H}{\partial p_i},\qquad
\frac{dp_i}{dt}=-\frac{\partial H}{\partial x^i},
\label{1}
\end{equation}
where $H$ (the Hamiltonian) is interpreted as the kinetic
energy:
$$
H=\frac{1}{2}\sum_{i,j=1}^n g^{ij}p_i p_j = \frac{1}{2}\sum_{i,j=1}^n
g_{ij}\dot x^i \dot x^j.
$$
Here $g^{ij}$ are the coefficients of the tensor inverse to the metric.

This system of equations is Hamiltonian on the
cotangent bundle $T^*M$ (with the standard symplectic form $\omega=\sum
dp_i\wedge dx^i$) and is called {\it the geodesic flow} of the Riemannian manifold
$(M,g)$.  Speaking more precisely, the geodesic flow is the
one-parameter group of diffeomorphisms defined by this system of
differential equations.

\subsection{Integrable geodesic flows}

We will be interested in the global behavior of geodesics on closed
Riemannian manifolds. Namely we want to distinguish the case of the
so-called integrable geodesic flows. Before giving the formal
definition, we consider some examples.

First take the two-dimensional sphere. The geodesics on it are
well known: these are equators. All of them are closed and of
the same length. Thus, the dynamics of the geodesic flow is very simple.
The second example are geodesics on the surface of revolution.
In a typical case the behavior of a geodesic is as follows: it moves on
the surface around the axis of revolution and at the same time
oscillates along this axis. If the geodesic is not closed, then its
closure is an annulus-like region on the surface.  As we see, the dynamics is
quite regular and is a superposition of rotation and oscillation.
An analogous picture can be seen on the three-axial ellipsoid.
This case is more complicated. Here there are two kinds of annulus-like
regions inside which geodesics move. But in the whole,
the behavior of geodesics is still regular.
Now consider an arbitrary closed surface without any special
symmetries (for example, deform a little bit the standard sphere
We shall see that the behavior of geodesics lose regularity and
becomes chaotic.

Regular behaviour is the characteristic property of integrable
geodesic flows which are formally defined as follows.

\begin{definition} {\rm
The geodesic flow (1) is called {\it completely
Liouville integrable}, if it admits $n$ smooth functions
$f_1(x,p),\dots,f_n(x,p)$ satisfying three conditions:

\smallskip

1) $f_i(x,p)$ is an integral of the geodesic flow, i.~e., is constant
along each geodesic line $(x(t),p(t))$;

\smallskip

2) $f_1,\dots,f_n$ pairwise commute with respect to the standard Poisson
bracket on $T^*M$, i.e.,  $\{f_i,f_j\}= \sum_{\alpha}\left(\frac{\partial f_i}
{\partial x^\alpha}\frac{\partial f_j}{\partial p_\alpha}-
\frac{\partial f_j}{\partial x^\alpha}\frac{\partial f_i}{\partial p_\alpha}\right)=0$;

\smallskip

3) $f_1,\dots,f_n$ are functionally independent on $T^*M$.
}\end{definition}

\begin{remark} {\rm
The third condition needs to be commented. The functional
independence of the integrals can be meant in three
different senses. The differentials of $f_1,\dots,f_n$ must be
linearly independent:

a) on an open everywhere dense subset,

b) on an open everywhere dense subset of full measure,

c) everywhere except for a piece-wise smooth polyhedron.

Instead of these conditions one can assume all the functions to be real
analytic. Then it suffices to require their functional independence
at least at one point (we will call such a situation {\it analytic integrability}).
}\end{remark}

The regularity of dynamics in the case of integrability
follows from the following classical Liouville theorem (see \cite{Ar}):

\begin{theorem} Let $X^n= \{f_1= c_1,\dots, f_n=c_n\}$ be a common level
surface of the first integrals of a Hamiltonian system.
If this surface is regular (i.e., the differentials of
$f_1,\dots,f_n$ are independent on it), compact and connected,
then

1) $X^n$ is diffeomorphic to the  $n$-dimensional torus;

2) the dynamics on this torus is quasi-periodic, i.e., can be linearized in appropriate angle
coordinates $\varphi_1,\dots,\varphi_n$:
$$
\varphi_1(t)= \omega_1 t,\ldots , \varphi_n(t)=\omega_n t.
$$
\end{theorem}

Thus, except a certain singular set, the phase space of the system turns
out to be foliated into invariant tori with quasi-periodic dynamics.
However the dynamics on the singular set may be rather
complicated (see below).

\subsection{Statement of the problem}

The general question discussed below can be formulated as follows:
Which closed smooth manifolds admit Riemannian metrics with integrable geodesic flows?
In other words, we want to divide all manifolds into two classes depending
on the fact if there exist or not integrable geodesic flows on them.
This question is rather complicated and so far we cannot expect any
complete answer.
At present there are, in essence, only two general approaches to the problem.

The positive answer to the existence question is strictly individual:
we can take a certain manifold (or a certain class of manifolds) and
construct on it an explicit example of a metric with integrable geodesic flow.
So far there is no other method except the explicit construction.
Thus the problem is reduced to studying new constructions of integrable
Hamiltonian systems in the particular case of geodesic flows.
The second approach deals with topological
obstructions to integrability. More precisely, the problem is
to find a topological property of a manifold which {\it is not compatible}
with integrability. Then all the manifold having such a
property belong to the second class, i.e., admit no integrable geodesic flows.

Notice that the character of first integrals is very important.
Usually one consider the integrals of three different types:
$C^\infty$-smooth, analytic or polynomials in momenta.
In the last case on can
fix or bound their degrees.
Besides, sometimes one has to require some topological restrictions for
the structure of the singular set.
Each time we deal with a specific problem and obtain a new
result, the most important of which we shall try to mention below.

Let us emphasize that in our paper we deal with topological obstructions
to integrability only. This means that we are mostly interested in the
principal possibility of constructing integrable systems on a given
manifold. There is another problem in Hamiltonian dynamics very natural
and important, namely,  the problem of analytical obstructions to
integrability, which can be formulated in a very general way as
follows.  Given a Hamiltonian function $H$, is the corresponding
Hamiltonian system integrable or not?  There are many quite powerful
methods to answer this question (see Kozlov \cite{K, K2}).
One of the most famous and classical is,
for example, the Painlev\'e-Kovalevskaya test.  Among modern
approaches to this problem one should mention the papers by Ziglin \cite{Zi},
Yoshida \cite{Yo1, Yo2}, and Ruiz and  Ramis (e.g., see \cite{Ru}).

Speaking of examples of integrable geodesic flows, in our paper we will
confine ourselves to the existence problem. However the complete description
of integrable geodesic flows of a certain type is extremely interesting and deep
problem. The algebro-geometric approach to the classification of
integrable geodesic flows has been developed by Adler and Van Moerbeke \cite{AM, AM2}
(case of $SO(4)$) and Haine \cite{Ha} (case of $SO(n)$).

\section{Classical examples of integrable geodesic flows}

\paragraph{Geodesic flows on two--dimensional surfaces.}
Classical examples of surfaces with integrable geodesic flows
are the two-dime\-nsional sphere
$S^2=\{x_1^2+x_2^2+x_3^2=1\}$ of constant curvature and the flat
torus. Geodesics on the sphere are equators (i.e., sections by the planes
passing through its center).
Since all geodesics are closed, in this case there are three
independent first integrals instead of two. As a "basis" one can take
linear integrals corresponding to infinitesimal rotations of the
sphere.
For example, the integral corresponding to the rotation about the axis
$Ox_3$ has the form
$$
f_3 (x,p)=p(\xi_3(x)),
$$
where $p\in T^*_x S^2$, and
$\xi_3=\partial/\partial \varphi$ is the vector field related to the
standard spherical coordinate $\varphi$ on $S^2$:
$\xi_3(x_1,x_2,x_3)=(-x_2, x_1, 0)$.  In terms of the tangent bundle
these integrals admit the following natural representation as one vector
integral:
$$
F(x,\dot x)=(f_1, f_2, f_3)=[x, \dot x]
$$
where $x\in S^2\subset \Bbb R^3$ and $\dot x\in T_xS^2$ are considered as vectors of
three-dimensional Euclidean space $\Bbb R^3$. The geometrical meaning of
$F$ is clear: this is the vector orthogonal to the plane to which the
geodesic belongs. These three linear integrals do not commute, but as two
commuting integrals one can take any of them and the Hamiltonian of the
geodesic flow $H=\frac12(f_1^2+f_2^2+f_3^2)$.

In  the case of a flat metric $ds^2=dx_1^2+dx_2^2$ on the torus, the
geodesics are quasiperiodic windings $x_i(t)=c_i t$ $(i=1,2)$, where
$x_1\,\mod 2\pi$, $x_2\,\mod{2\pi}$ denote standard angle coordinates.
The first commuting integrals are the corresponding momenta $p_1$ and
$p_2$.
The Hamiltonian of the flow is expressed by $p_1$ and $p_2$ in the obvious way:
$H(x_1,x_2,p_1,p_2)=\frac 12(p_1^2 + p_2^2)$,
or more generally,
$H(x_1,x_2,p_1,p_2)=\frac 12(ap_1^2 +2bp_1p_2+ cp_2^2)$.

The next classical examples are metrics of revolution and
Liouville metrics.

\begin{theorem}\label{Clairaut}{\rm (Clairaut)}
The geodesic flow on a surface of revolution admits a
non-trivial linear integral and, consequently, is integrable.
\end{theorem}

The Clairaut integral has the same form as $f_3$ in the case of the
sphere (if we consider a surface of revolution about the axis $Ox_3$).
It admits also the following geometrical description.
Let $\psi$ be the angle between the geodesic and the parallel on the
surface of revolution, $r$ be the distance to the revolution axis. Then
the Clairaut integral is $r\cos\psi$.

The existence of such an integral is the reflection of the
following classical result (Noether theorem):  let a
Riemannian metric $g_{ij}$ admit a one-parameter isometry group
$\phi^s:  M\to M$. Then the corresponding geodesic flow has the linear
integral of the form $f_\xi(x,p)=p(\xi(x))$, where
$\xi(x)=\frac{d}{ds}|_{s=0} \phi^s(x)$ is the vector field
associated with the one-parameter group $\phi^s$ (that is, the
corresponding infinitesimal isometry). More generally:
if a metric admits an isometry group $G$, then the corresponding geodesic
flow has an algebra of first integrals which is isomorphic to the Lie
algebra of the group $G$.

\begin{theorem}\label{Liouville} {\rm (Liouville)}
The geodesic flow of the metric
\begin{equation}
ds^2= \bigl( f(x_1)+g(x_2) \bigr) (dx_1^2+dx_2^2)
\label{2}
\end{equation}
admits the non-trivial quadratic integral of the form
\begin{equation}
F(x,p)=\frac{g(x_2)p_1^2-f(x_1)p_2^2}{f(x_1)+g(x_2)}
\label{3}
\end{equation}
and, consequently, is integrable.
\end{theorem}

This formula, in general, is local. It is not hard to construct an
example of a metric on a two-dimensional surface, which admits
representation (\ref{2}) in some local coordinates at each point, but on the whole surface, the
integrals given by (\ref{3}) cannot be arranged in one globally defined integral.

One of such examples is the constant negative curvature metric on
a surface of genus $g>1$.  Locally its geodesic flow admits a
quadratic integral (speaking more precisely, there are three
independent linear integrals from which one can combine a
quadratic one). However, there are no such  global integrals
(see the section below).

\paragraph{Projectively equivalent metrics.}
There is another interesting class of manifolds closely related to our
problem. These are manifolds admitting
two {\it projectively} (or {\it geodesically}) equivalent
Riemannian metrics $g$ and $\tilde g$, that is, metrics having the same
geodesics (considered as unparametrized curves) (see Levi-Civita \cite{LC}).
If $g$ and $\tilde g$ are in general position, i.e., there exists at least one point
at which the operator $g\tilde g^{-1}$ has simple spectrum, then the
geodesic flows of the both metrics are integrable. Moreover, the first
integrals are all quadratic or linear functions. A rather elegant proof of
this fact has been obtained by V.Matveev and P.Topalov \cite{MT, To}
(it is interesting that the corresponding Hamilton-Jacobi equations
admit separation of variables and that the problem can be
considered from a bi-Hamiltonian point of view,  see \cite{BM}).
The open problem is to describe the class of all such manifolds.

\paragraph{Geodesics on the ellipsoid.}
Finally, as one of the most beautiful examples we have
the geodesic flow of the standard metric $ds^2=\langle dx,dx\rangle$
on the $(n-1)$--dimensional ellipsoid \cite{Mo1, Mo2, Kn, Fe}
$$
Q=\{\langle x,A^{-1}x\rangle =1\}\subset \mathbb{R}^{n},
\quad A={\mathrm{diag}}(a_1,\dots,a_n), \quad a_1>a_2>\dots>a_n.
$$
The problem was solved by Jacobi by separation of variables in
elliptic coordinates \cite{Ja}. Moser, in his famous papers \cite{Mo1, Mo2},  found an L-A pair representation and commuting integrals in Euclidian coordinates by
using geometry of quadrics.

\begin{theorem}\label{Moser} {\rm (Moser, 1980)}
The functions $F_k$:
$$
F_k(x,\dot x)=\dot x_k^2+
\sum_{l\ne k}\frac{(x_k \dot x_l-x_l \dot x_k)^2}{a_k-a_l}, \quad k=1,\dots,n,
$$
restricted to the tangent bundle $TQ\{x,\dot x\}$
Poisson commute and give complete integrability of the geodesic flow
on the ellipsoid $Q$.
\end{theorem}

The geometric interpretation  of the integrability is described by the
Chasles theorem: the tangent line of a geodesic $x(t)$
on $Q$ is also tangent to a fixed set of confocal quadrics
$Q(\alpha_i)= \{\langle (A-\alpha_iId)^{-1}x,x\rangle=1\}$,
$\alpha_i\in {\mathbb R}$, $i=1,\dots,n-2$ (for example, see
\cite{Kn, Fe}).

\section{Topological obstructions to integrability}

\subsection{Case of two-dimensional surfaces}

The first results on topological obstructions to integrability
relate to the case of two-dimensional surfaces.

 \begin {theorem} {\rm (Kozlov, 1979 \cite{Ko})}
Two-dimensional surfaces of genus $g>1$ admit no analytically
integrable geodesic flows.
\end{theorem}

The original proof essentially  used some delicate
properties of analytic functions. But later it was
understood that the analyticity condition could be essentially
weakened (see Taimanov \cite{T1, T2}).
However, it is still not clear if it is possible to omit any
additional conditions to the first integrals.

\begin{question}

Given a two-dimensional surface of genus $g>1$,  do there exist integrable
geodesic flows on it with $C^\infty$-smooth integrals?
\end{question}

Here is another result related to the same class of surfaces.

\begin {theorem} {\rm (Kolokol'tsov, 1982 \cite{Ko})} \label{Kolokoltsov}
Two-dimensional surfaces of genus $g>1$ admit no geodesic flows
integrable by means of an integral polynomial in momenta (the
coefficients of this polynomial are assumed to be smooth functions without
any analyticity conditions).
\end{theorem}

The idea of the proof of Theorem \ref{Kolokoltsov} is rather different from that of
Kozlov's theorem. By using the polynomial integral one constructs a
certain holomorphic form on the given surface. Then analyzing
its zeros and poles one can estimate the genus of the surface.

\begin{question}
Is there a multidimensional analog of Kolokol'tsov theorem? In
other words, are there topological obstructions to polynomial
integrability?
\end{question}

If we confine ourselves to linear and quadratic integrals, then the problem
is getting simpler and it is, probably, possible to obtain the complete
list of manifolds with linearly and quadratically integrable geodesic flows. The point
is that under some additional conditions such flows admit separation of
variables on the configuration space. These variables will have, however,
certain singularities and the problem is reduced to studying the topology
of manifolds admitting global coordinate systems with special types of
singularities. Such an approach has been used by Kiyohara \cite{Ki1},
but the final answer is not yet obtained.

\subsection{Topological obstructions in the case of non simply connected manifolds}

The next fundamental result is due to I. Taimanov \cite{T1, T2}.

\begin{theorem} {\rm (Taimanov, 1987)} \label{Taimanov}
If a geodesic flow on a closed manifold $M$ is analytically
integrable, then

1) the fundamental group of $M$ is almost commutative (i.e., contains
a commutative subgroup of finite index);

2) if $\dim H_1(M,\Bbb Q)=d$, then $H^*(M,\Bbb Q)$ contains a subring
isomorphic to the rational cohomology ring of the $d$-dimensional torus;

3) if $\dim H_1(M,\Bbb Q)=n=\dim M$, then the rational cohomology
rings of $M$ and  of the $n$-dimensional torus are isomorphic.
\end{theorem}

The idea of the proof is purely topological and the analyticity
condition is not essential. In fact, I.Taimanov proved this result
under the much weaker assumption that a geodesic flow is geometrically simple.
This means that the structure of the singular set where the first
integrals are dependent is not too complicated from the topological point
of view. Speaking more precisely,
to this singular set one should add some new "cuts"
in such a way that the rest becomes a trivial fibration into
Liouville tori over a disjoint union of discs.  Besides the geometric
simplicity condition takes into account some properties of the projection
of this "completed" singular set from the cotangent bundle onto the
configuration space (see \cite{T1, T2}).

\subsection{Topological entropy and integrability of geodesic flows}

In 1991 Paternain suggested an approach to finding topological
obstructions to integrability of geodesic flows based on the
notion of {\it topological entropy} \cite{P1, P2}.
The topological entropy is a characteristic of a dynamical
system on a compact manifold, which measures, in a certain sense, its
chaoticity.
Since, as a rule, integrable Hamiltonian systems have zero
topological entropy, one can proceed as follows. First one may
try to estimate the topological entropy of a geodesic flow on a
given manifold by purely topological means.
Very often one can do that even without any information about the
riemannian metric: if the topology of a manifold is sufficiently
complicated (see examples below), then the topological entropy of any
geodesic flow has automatically to be positive. The second part of the
problem is to prove that the integrability of a geodesic flow implies
indeed vanishing of the topological entropy
(perhaps under some additional conditions to the first integrals).

Recall the definition of topological entropy.
Let $F^t$ be a dynamical system on a compact manifold $X$
considered as a one-parameter group of diffeomorphisms.
Suppose that we want to approximate this systems up to $\varepsilon >0$ on
a segment $[0,T]$ by using only finite number of solutions. In other
words, we want to choose a finite number of points
$x_1,\dots,x_{N(\varepsilon,T)}$ in such a way that for any other point
$y\in X$ there exists $x_i$ satisfying
$$
\dist (F^t(y),F^t(x_i)) < \varepsilon
$$
for any $t\in[0,T]$.
Here  $\dist$ denotes any metric compatible with the
topology of $X$. Suppose now that $N(\varepsilon ,T)$ is the minimal
number of such points $x_i$ and consider its asymptotics as
$\varepsilon\to 0$ and $T\to\infty$.

\begin{definition}{\rm
The topological entropy of the flow $F^t$ is define to be
$$
h_{top} (F^t)=\lim_{\varepsilon\to 0} \limsup_{T\to \infty} \frac{\ln
N(\varepsilon, T)}{T}.
$$
}\end{definition}

The next theorem is the first result showing how the topology of a
manifold affect the topological entropy of geodesic flows.

\begin{theorem} {\rm (Dinaburg, 1971 \cite{Di})}
If the fundamental group $\pi_1(M)$ of $M$ has exponential growth,
then the topological entropy of the geodesic flow is positive for any
smooth Riemannian metric on $M$.
In particular, the topological entropy of any geodesic flow on a
two-dimensional surface of genus $g>1$ is positive.
\end{theorem}

Recall that the topological entropy of a geodesic flow is
that of its restriction onto a compact isoenergy surface
$Q^{2n-1}=\{ H(x,p)=1\}\subset
T^*M^n$, $H(x,p)=\frac{1}{2}|p|^2$.

It is important that the entropy approach works successfully in the
case of simply-connected manifolds when Kozlov's and Taimanov's
theorems cannot be applied.

\begin{theorem} {\rm (Paternain, 1991 \cite{P1, P2})}
If a simply-connected manifold is not rationally elliptic, then the
topological entropy of any geodesic flow on it is positive.
\end{theorem}

The rational ellipticity means that the rational homotopy
groups  of $M$ are trivial starting from a certain
dimension $N$, i.e., $\pi_n(M)\otimes\Bbb Q=0$ for any $n>N$. The next
result not only guarantees the positiveness of the topological entropy for
a large class of manifolds, but also allows one to estimate it from below.

\begin{theorem} {\rm (Babenko, 1997 \cite{Bab})} The topological entropy of a geodesic flow on a simply
connected Riemannian manifold $(M,g)$ admits the following estimate
$$
h_{top}(g)\ge D^h(M,g)^{-1}\limsup_{n\to\infty}\frac{\ln(\rk {\pi_n(M)})}{n},
$$
where $D^h(M,g)$ is the homology diameter of $(M,g)$.
\end{theorem}

Note that the limit $\limsup_{n\to\infty}\frac{\ln(\rk
{\pi_n(M)})}{n}$ is equal to zero only
for rationally elliptic manifolds, otherwise this is always a positive number. The
homology diameter of a manifold depend on the choice of
$g$ and is, therefore, a geometric characteristic of a manifold
(exact definition can be found in \cite{Bab}).

To use the topological entropy as an obstruction to integrability, it
was necessary to show vanishing topological entropy for integrable
geodesic flows. Under some rather strong additional assumptions this
statement holds indeed.

\begin{theorem} {\rm (Paternain, 1991 \cite{P1, P2})}
Let $(M,g)$ be a smooth compact Riemannian manifold. Suppose that the
geodesic flow on it is integrable in the class of non-degenerate first
integrals. Then the topological entropy of this flow vanishes.
\end{theorem}

The non-degeneracy of integrals means the following. First
consider the case when $x$ is a singular point for all the first integrals of the system.
Then the non-degeneracy is equivalent to the existence of such an integral $f$ that
$x$ is non-degenerate in the usual sense of the Morse theory, that is,
$\det\left( \frac{\partial^2 f}{\partial x^i\partial x^j} \right)\ne 0$.
If some integrals have non-zero differentials at $x$, then we make
local symplectic reduction by the action of these integrals
after which we get to the previous situation. The point
$x$ is called non-degenerate for the initial system if it is
such for the reduced one.
The integrals of a Hamiltonian system are called non-degenerate
if every point $x$ of a symplectic manifold is non-degenerate.

However, the non-degeneracy assumption is rather strong
and, in the multidimensional case, holds very rarely.
For example, the degeneracy
appears in such a natural case as the geodesic flow on the $n$-dimensional
ellipsoid ($n\ge 3$).
One of the reasons is the existence of stable degenerate singularities
which cannot be avoided by small perturbation and, consequently, are generic.

\section{Counterexamples}

It has been, however, understood recently that additional assumptions on the first
integrals cannot be completely omitted.
In other words, in general case
neither topological entropy, nor "complexity" of the
fundamental group is an obstruction to integrability of geodesic flows.

The first interesting example of an "exotic" integrable flow was constructed by L. Butler \cite{Bu}.

\begin{theorem}\label{Butler}
{\rm (Butler, 1998)} There is a three-dimensional Riemannian
(real-analytic) NIL-manifold $(M^3, g)$ such that

1)  the geodesic flow of the metric $g$ on $M^3$ is completely integrable by means of
$C^\infty$-smooth first integrals (moreover, two of these integrals are real-analytic functions);

2)  the fundamental group $\pi_1(M)$ is not almost commutative and has polynomial growth;

3)  the topological entropy of the geodesic flow vanishes.
\end{theorem}

This example shows that in the smooth case the statement analogous to
Taimanov's theorem (Theorem \ref{Taimanov}) fails.
Thus, the geometric simplicity assumption introduced by Taimanov
is really very important (Butler's example is not geometrically simple).
Besides this is the first example of the
situation when a real-analytic flow is integrable, but its integrals
cannot be real-analytic functions.

The topological structure of $M^3$ in Butler's example is quite simple.
This is a fibration
$M^3 {{\quad T^2}\over{\quad}}\!\!\!\!\!{\longrightarrow} S^1$
over the circle with the torus as a
fiber and with the monodromy matrix of the form
$$
A=\left(\begin{array}{cc}
 1 & 1 \\ 0 & 1
\end{array}\right)
$$
In other words, to reconstruct the manifold one needs to take the direct product  $T^2 \times [0,2\pi]$ and then to glue its feet
$T^2\times\{0\}$ and $T^2\times \{2\pi\}$ by the linear map given by the matrix $A$ in angle coordinates.

Using the idea of Butler, a year later the first author and Taimanov
in \cite{BT1} constructed an example of a three-dimensional manifold with
integrable geodesic flow having positive topological entropy.

\begin{theorem}\label{Bol_Tai}
{\rm (Bolsinov, Taimanov, 1999)} There is a three-dimensional Riemannian
(real-analytic) SOL-manifold $(M^3, g)$ such that

1)  the geodesic flow of the metric $g$ on $M^3$ is completely
integrable by means of $C^\infty$-smooth first
integrals (moreover, two of these integrals are real-analytic functions);

2)  the fundamental group $\pi_1(M)$ is not almost commutative and has exponential growth;

3)  the topological entropy of the geodesic flow is positive.
\end{theorem}

In this example, $M^3$ has "almost the same" structure as in
Butler's example: it suffices to replace the matrix $A$
by an integer hyperbolic matrix, for instance,
$$
B=\left(\begin{array}{cc}
 2 & 1 \\ 1 & 1
\end{array}\right).
$$

The Riemannian metric $g$ on $M^3$
is described as follows. Let $x$, $y$, $z$
be local coordinates on $M$, where $z\mod 2\pi$ is an angle coordinate on
the base $S^1$, and $x\mod 2\pi, y\mod 2\pi$ are angle coordinates on
the fiber  $T^2$. Then
$$
ds^2 = d\tilde s_z^2 + dz^2, $$ where $$ d\tilde s_z^2=a_{11}(z)dx^2 + 2a_{12}(z)dxdy +
a_{22}(z)dy^2
$$
is a flat metric on the fiber $T_z^2$
over $z\in S^1$.  The coefficients $a_{11}, a_{12}, a_{22}$
are chosen so that the metric turns out to be smooth on the whole manifold $M$.

Since the coefficients of the metric do not depend on $x$ and
$y$, the corresponding momenta $p_x$, $p_y$ are commuting
first integrals of the geodesic flow. The only problem is that
$p_x$ and $p_y$ are not globally defined on $M^3$ because of
non-triviality of the monodromy of the $T^2$-fibration. However, it is
possible to construct other functions $F_1=F_1(p_x,p_y)$ and
$F_2=F_2(p_x,p_y)$ which will be preserved under the monodromy action.  As
such functions one should take invariants of the action of the
cyclic group $\Bbb Z$ generated by the linear transformation
${B^{-1}}^\top$ on the two-dimensional space
$\Bbb R^2(p_x, p_y)$. It is interesting to remark that the orbit structure
of this action is such that only one of the invariants is an analytic
function. The second can be chosen  $C^\infty$-smooth, but not real-analytic.
The functions $F_1$ and $F_2$ (together with the Hamiltonian) guarantee
the complete integrability of the geodesic flow on $(M,g)$.

The positiveness of the topological entropy easily follows
from Dinaburg's theorem, but also can be explained
directly. The point is that the geodesic flow admits a
natural invariant manifold $N$ formed by the geodesics of the form
$x(t)=\const, y(t)=\const, z(t)=t$.
It is easily seen that the union of such geodesics is a submanifold
in $T^*M$ diffeomorphic to the base $M$.
The geodesic flow $\phi^t$ restricted onto $N$ preserves the
$T^2$--fibers. Moreover the  $2\pi$-shift along the flow transforms each
fiber into itself $\phi^{t=2\pi}: T^2 \to T^2$ and coincides with
the hyperbolic automorphism of the torus given by the matrix $B$.
It is well-known (as one of basic examples) that the topological entropy
of such an automorphism is positive and equals to $\ln\lambda$, where
$\lambda$ is the maximal eigenvalue of $B$ (see, for example, \cite{KH}).
Thus the geodesic flow admits a subsystem with positive
topological entropy and, consequently, possesses this property itself.

The constructions of Theorems \ref{Butler}, \ref{Bol_Tai} are naturally generalized to
the case of arbitrary dimension \cite{BT2, Bu2}.

Note that the topological structure of the singular set in these examples
is quite simple. This is a finite polyhedron whose strata are not just
only smooth but also real-analytic submanifolds.
"Non-analytic" is the way of how Liouville tori approach this singular set.

It is worth to explain why the above geodesic flows are not
geometrically simple in the sense of Taimanov \cite{T1, T2}.
The point is that
the base of the foliation into Liouville tori is not simply
connected. To make it such (as required in the definition of geometric
simplicity) we need additional cuts of the base. But this is impossible to
do by a "geometrically simple" way: in each tangent space such a cut
will consist of infinitely many two-dimensional planes.

The above examples lead to the following questions:

\begin{question}
Which additional properties of the first integrals garantee vanishing the
topological entropy?
\end{question}

\begin{question}

Is analyticity such a condition? In other words, is the topological entropy an obstruction to analytic
integrability?
\end{question}

\begin{question}
Is it possible to include an arbitrary (as chaotic as one wishes)
dynamical system as a subsystem into an integrable Hamiltonian system of
higher dimension?
\end{question}

\section{Geodesic flows on homogeneous spaces and bi-quotients of Lie groups}

Since the integrability is closely related to the existence of some symmetries (possibly hidden),
for the construction of
multi-dimensional integrable examples, as a rule we should use
metrics with large symmetry groups. In the final construction
the symmetry can be removed by algebraic modification of metrics.

The classical example is the geodesic flow of a left-invariant metric on
the Lie group $SO(3)$. The geodesic flow of such a metric
describes the motion of a rigid body
about a fixed point under its own inertia.
This problem was solved by Euler.
In general, the geodesic flow of a left-invariant metrics on a Lie group $G$
after $G$--reduction reduces to the {\it Euler equations} on $\mathfrak g^*$
(the dual space of the Lie algebra $\mathfrak g$), which are Hamiltonian equations
with respect to the Lie-Poisson bracket on $\mathfrak g^*$ \cite{Ar}.

A multidimensional generalization of the Euler case has been suggested by
Manakov \cite{Ma}. Using his idea, Mishchenko and Fomenko proposed the argument
shift method (see below) and constructed integrable examples
of Euler equations for all compact groups \cite{MF1} and
proved the integrability of the original geodesic flows \cite{MF2, MF3}.

\begin{theorem}
{\rm (Mishchenko, Fomenko 1976)}
Every compact Lie group admits a family of left-invariant
metrics with
completely integrable geodesic flows.
\end{theorem}

There are many other important constructions on various Lie algebras
(we mention just some of the review papers and books
\cite{BCK, BB, F, FT, RS, Pe}).
In particular, the problem of algebraic
integrability of geodesic flows on
$SO(4)$ and $SO(n)$ is studied  in  \cite{AM, AM2, Ha}.

Throughout the paper, by a {\it normal} $G$--invariant Riemannian metric
on the homogeneous space $G/H$ of a compact group $G$,
we mean the metric induced from a bi-invatiant metric on $G$.

\begin{theorem}
{\rm (Thimm 1981, Mishchenko 1982)}
Geodesic flows of normal metrics
on compact symmetric spaces $G/H$ are completely integrable.
\end{theorem}

These results are generalized by Brailov, Guillemin and Sternberg and
Mikityuk. Brailov applied the Mishchenko--Fomenko construction
to the symmetric spaces and obtained families of non $G$--invariant
metrics with integrable flows on symmetric spaces \cite{Br2, Br3}.
Guillemin and Sternberg \cite{GS1, GS2} and Mikityuk \cite{Mik3}
described the class of homogeneous spaces $G/H$ on which
all $G$--invariant Hamiltonian systems are integrable by
means of Noether integrals \cite{GS1, GS2, Mik3}.
It appears that in those cases $(G,H)$ is a spherical (or Gelfand) pair.

If $G/H$ is a symmetric space then $(G,H)$ is a spherical pair,
but there exist spherical
pairs which are not symmetric.
Note that, for $G$ compact, $(G,H)$ is a spherical
pair if and only if
$G/H$ is a weekly symmetric space (see \cite{Vi} and references therein).

Examples of homogeneous, non (weekly) symmetric spaces $G/H$
with integrable geodesic flows are given by Thimm \cite{Th}, Paternain and  Spatzier \cite{PS}
and Mikityuk and Stepin \cite{MS}. Also,
to this list now we can add the above mentioned examples
constructed by Butler \cite{Bu} and Taimanov and the first author \cite{BT1, BT2}.

Particular examples of non-homogeneous manifolds
(bi-quotients of compact Lie groups) with integrable geodesic flows are obtained
by Paternain and  Spatzier \cite{PS} and Bazaikin \cite{Ba}.

It appears that many of those results can be considered
together within the framework of non-commutative integrability.
This approach allowed us to obtain general theorems on integrability
of geodesic flows on homogeneous spaces and bi-quotients of
compact Lie groups \cite{BJ1, BJ2}.

\subsection{Non-commutative integrability}

There are a lot of examples of integrable Hamiltonian systems with $n$ degrees of freedom that
admit more than $n$ (noncommuting) integrals.
Then $n$ dimensional Lagrangian tori are foliated by
lower dimensional isotropic tori
(sometimes such systems are called {\it superintegrable}).
The concept of non-commutative integrability
has been introduced by Mishchenko and Fomenko \cite{MF2, MF3}
(see also \cite{N, Br1, F}).

Let $M$ be a $2n$--dimensional symplectic manifold.
Let $(\mathcal F,\{\cdot,\cdot\})$ be a Poisson subalgebra
of  $(C^\infty(M),\{\cdot,\cdot\})$. Suppose that in the neighborhood of a generic point $x$
we can find exactly $l$ independent functions $f_1,\dots,f_l \in \mathcal F$
and the corank of the matrix $\{f_i,f_j\}$ is equal to some constant $r$.
Then numbers $l$ and $r$ are called {\it differential dimension} and
{\it differential index} of $\mathcal F$ and they are denoted by
$\ddim\mathcal F$ and $\dind\mathcal F$, respectively.
The algebra $\mathcal F$ is called {\it complete} if:
$$
\ddim{\mathcal F}+\dind{\mathcal F}=\dim M.
$$

In fact, this definition is equivalent to any of the following three conditions.

Take a generic point $x$ and $l$ independent functions
$f_1,\dots,f_l\in \mathcal F$ in some small neighborhood $U$ of $x$. Then

\smallskip
(i) Common level sets of  $f_1,\dots,f_l$ in $U$ are isotropic.

\smallskip
(ii) The subspace generated by $df_1(x),\dots,df_l(x)$ is
coisotropic in $T^*_x M$.

\smallskip
(iii) If there is a function $f$ which
commute with $\mathcal F$ on $U$ then $f=f(f_1,\dots,f_l)$ on $U$.

\smallskip

In particular, we shall say that
the algebra $\mathcal F$ {\it is complete at} $x$
if one of the conditions (i--iii) is satisfied.

The Hamiltonian system  $\dot x=X_H(x)$
is {\it completely integrable in the non-commutative sense} if it
possesses a complete algebra of first integrals $\mathcal F$.
Then  each connected compact component of a regular level set of the
functions $f_1,\dots,f_l\in\mathcal F$
is an $r$--dimensional invariant isotropic torus $\mathbb{T}^r$
(see \cite{MF2, N, MF3, Br1, F}).
Similarly as in the Liouville theorem,
in a neighborhood of $\mathbb{T}^r$ there are {\it generalized action-angle
variables} $p,q,I,\varphi$, defined in a toroidal domain
$\mathcal O=\mathbb{T}^r\{\varphi\}\times B_\sigma\{I,p,q\}$,
$$
B_\sigma=\{(I_1,\dots,I_r,p_1,\dots,p_k,q_1,\dots,q_k)\in\mathbb{R}^l,\;
\sum_{i=1}^r I_i^2 + \sum_{j=1}^k q_j^2+p_j^2 \le \sigma^2\}
$$
such that the symplectic form becomes
$
\omega=\sum_{i=1}^r dI_i\wedge d\varphi_i+\sum_{i=1}^k dp_i\wedge dq_i,
$
and the Hamiltonian function depends only on $I_1,\dots,I_r$.
The Hamiltonian equations take the following form in action-angle coordinates:
$$
\dot \varphi_1=\omega_1(I)=\frac{\partial H}{\partial I_1},
\dots, \dot \varphi_r=\omega_r(I)=\frac{\partial H}{\partial I_r},
\quad \dot I=\dot p=\dot q=0.
$$

\begin{example}{\rm
Consider a Riemannian manifold $(Q,ds^2)$ with closed geodesics:
for every $x\in Q$,
all geodesics starting from $x$ return back to the same point.
Then one can find a complete algebra of integrals $\mathcal F$
with $\dind\mathcal F=1$.
In other words,
the geodesic flow is completely integrable in non-commutative sense \cite{BJ2}.
}\end{example}

Note that the concept of noncommutative integrability
can be naturally extended to Poisson manifolds $(N,\{\cdot,\cdot\})$.
The algebra ${\mathcal F}$ is {\it complete} if
$$
\ddim{\mathcal F}+\dind{\mathcal F}=\dim N+\corank\{\cdot,\cdot\},
$$
i.e., the restriction of $\mathcal F$ to a generic symplectic leaf $M$ of $N$
is complete on $M$.
Also, if a Hamiltonian system $\dot x=X_H(x)$ on $N$
possesses a complete algebra of first integrals $\mathcal F$,
then (under compactness condition) $N$ is almost everywhere foliated by
$(\dind{\mathcal F}-\corank\{\cdot,\cdot\}$)-dimensional
invariant tori with quasi-periodic dynamics.

\subsection{Integrable geodesic flows on $G/H$ and $K\backslash G/H$}

Let a connected compact Lie group $G$ act on a $2n$--dimensional connected
symplectic manifold $(M,\omega)$. Suppose the action is Hamiltonian, i.e.,
$G$ acts on $M$ by  symplectomorphisms and there is a well-defined momentum mapping
$\Phi: M\to \mathfrak g^*$ ($\mathfrak g^*$ is the dual space of the Lie algebra $\mathfrak g$)
such that one-parameter subgroups of symplectomorphisms  are generated by the
Hamiltonian vector fields of functions
$f_\xi(x)=\Phi(x)(\xi)$, $\xi\in\mathfrak g$, $x\in M$  and
$f_{[\xi_1,\xi_2]}=\{f_{\xi_1},f_{\xi_2}\}$.
Then $\Phi$ is  equivariant with respect to the given action of  $G$ on $M$
and the co-adjoint action of $\mathfrak G$ on $\mathfrak g^*$.
In particular, if $\mu\in\Phi(M)$ then the co-adjoint
orbit ${\mathcal O}_G(\mu)$ belongs to $\Phi(M)$.

Consider the following two natural classes of functions on $M$.
Let ${\mathcal F}_1$ be the set of functions in
$C^\infty(M)$ obtained by pulling-back the algebra $C^\infty(\mathfrak g^*)$
by the moment map ${\mathcal F}_1=\Phi^*C^\infty(\mathfrak g^*)$.
Let ${\mathcal F}_2$ be  the set of $G$--invariant functions in $C^\infty(M)$.
The mapping $h\mapsto f_h=h\circ\Phi$ is a morphism of Poisson structures:
$$
\{f_{h_1}(x),f_{h_2}(x)\}=
\{h_1(\mu),h_2(\mu)\}_{\mathfrak g^*}, \quad \mu=\Phi(x),
$$
where $\{\cdot,\cdot\}_{\mathfrak g^*}$ is the Lie-Poisson bracket on
$\mathfrak g^*$:
$$
\{f(\mu),g(\mu)\}_{\mathfrak g^*}=\mu([df(\mu),dg(\mu)]),
\quad f,g: \mathfrak g^*\to \Bbb{R}.
$$
Thus, ${\mathcal F}_1$ is closed under the Poisson bracket.
Since $G$ acts in a Hamiltonian way,
${\mathcal F}_2$ is closed under the Poisson bracket as well.
In other words, ${\mathcal F}_1$ and ${\mathcal F}_2$ are
Lie subalgebras in $C^\infty(M)$.

The second essential fact is that $h\circ\Phi$
commute with any $G$--invariant function
(the Noether theorem).
In other words:
$\{{\mathcal F}_1,{\mathcal F}_2\}=0$.

The following theorem, although it is a reformulation of
some well known facts about the momentum mapping (see \cite{GS3, LM}),
is fundamental in the considerations below. This
formulation was suggested by A. S. Mishchenko.

\begin{theorem} \label{Mis}
The algebra of functions ${\mathcal F}_1 + {\mathcal F}_2$ is complete:
$$
\ddim({\mathcal F}_1+ {\mathcal F}_2)+ \dind({\mathcal F}_1+{\mathcal F}_2)=\dim M.
$$
\end{theorem}

Let $A\subset C^\infty (\mathfrak g^*)$ be a Lie subalgebra.
By ${\mathcal A}$ we shall denote
the pull-back of $A$ by
the moment map:
${\mathcal A}=\Phi^*(A)=\{f_h=h\circ \Phi, \; h\in A\}$.
Let $\mathcal B$ be a subalgebra of $\mathcal{F}_2$.

\begin{corollary} {\rm \cite{BJ2}} \label{Bol_Jov}
(i) ${\mathcal A}+{\mathcal F}_2$ is a complete algebra on $M$ if and only if

$A$ is a complete algebra on the generic orbit ${\mathcal O}_G(\mu)\subset \Phi(M)$
of the co-adjoint action.

(ii) If $\mathcal B$ is complete (commutative) subalgebra and
$A$ is complete (commutative) algebra
on the orbit ${\mathcal O}_G(\mu)$, for generic $\mu \in \phi(M)$
then ${\mathcal A}+{\mathcal B}$
is complete (commutative) algebra on $M$.
\end{corollary}

Now, let $M$ be the cotangent bundle $T^*(G/H)$ with the natural
$G$ action. By $ds^2_0$ we shall denote the normal $G$--invariant
Riemannian metric on the homogeneous space $G/H$, induced by a bi-invariant metric on  $G$.

The Hamiltonian of the metric $ds^2_0$ Poisson commute with
$\mathcal{F}_1$ and $\mathcal{F}_2$. Thus, from Theorem \ref{Mis}
we get

\begin{theorem} {\rm \cite{BJ1}}
Let $G$ be a compact Lie group.
The geodesic flows of normal metrics $ds^2_0$ on the
homogeneous spaces $G/H$ are completely integrable (in non-commutative sense).
\end{theorem}

\begin{remark}{\rm
The dimension of invariant isotropic tori $T^*G/H$ is
$\dim pr_\mathfrak v(\ann_\mathfrak g(v))$
(see remark \ref{dimension} in the next section).}
\end{remark}

Note that for $\mathcal{F}_1$ and $\mathcal{F}_2$
we can take analytic functions, polynomial in momenta.
For example, in the case of Lie groups, these functions
are polynomials on $\mathfrak g^*$ shifted to $T^*G$ by right and left
translations, respectively.

A similar construction can be applied to bi-quotients of compact groups.
Consider a subgroup $U$ of $G\times G$ and define the action of $U$ on $G$ by:
$$
(g_1,g_2)\cdot g =g_1gg_2^{-1}, \quad (g_1,g_2)\in U,\; g\in G.
$$

If the action is free then the orbit space $G//U$ is a
smooth manifold called a {\it bi-quotient} of the Lie group $G$.
In particular, if $U=K\times H$, where
$K$ and $H$ are subgroups of $G$, then the bi-quotient $G//U$
is denoted by $K\backslash G/H$.
A bi-invariant metric on $G$, by submersion (see section \ref{int_red}),
induces a normal Riemannian metric $ds^2_0$ on $G//U$. Note that every bi-quotient $G//U$
is canonically isometric to $\Delta G\backslash G\times G/U$,
for a certain normal metric on $\Delta G\backslash G\times G/U$.
Here $\Delta G$ denotes the diagonal subgroup (see \cite{Wi}).

On $T^*(K\backslash G/H)$ there exist algebras $\mathcal{F}_1$ and $\mathcal{F}_2$ analogous to the above algebras on homogeneous spaces.
For $\mathcal{F}_1$ we take all polynomials on $\mathfrak g^*$ invariant with respect to the
$K$-action and extend them to right invariant functions
on $T^*G$. These functions give well defined functions on  $T^*(K\backslash G/H)$
since they are invariant with respect to the $K\times H$--action on $T^*G$.
Similarly, for $\mathcal{F}_2$ we take all polynomials
on $\mathfrak g^*$ invariant with respect to the $H$-action, extend them to left invariant functions
on $T^*G$ and consider as functions on $T^*(K\backslash G/H)$.
(When $K$ is trivial these algebras are precisely the algebras
described above). It is clear that in such a way we obtain integrals
of geodesic flow of $ds^2_0$.

\begin{theorem} {\rm \cite{BJ2}}
The algebra of functions ${\mathcal F}_1 + {\mathcal F}_2$ is complete:
$$
\ddim({\mathcal F}_1+ {\mathcal F}_2)+
\dind({\mathcal F}_1+{\mathcal F}_2)=\dim T^*(K\backslash G/H)
$$
and, therefore, the geodesic flows of  $ds^2_0$ on the
bi-quotient $ K\backslash G/H$  is completely integrable (in non-commutative sense).
\end{theorem}

\section{Complete commutative algebras on $T^*(G/H)$}

\subsection{Mishchenko--Fomenko conjecture}

Mishchenko and Fomenko stated the conjecture that  non-commutative
integrable systems $\dot x=X_H(x)$
are integrable in the usual commutative sense by means of
integrals that belong to the same functional class as the original
non-commutative algebra of integrals \cite{MF2, F}.
Note that, locally, non-commutative integrability
always implies commutative integrability.
For example, we can just take commuting functions
$\{I_1,\dots,I_r,p_1^2+q_1^2,\dots,p_k^2+q_k^2\}$.

When $\mathcal F=\Span_{\mathbb R}\{f_1,\dots,f_l\}$ is a
finite-dimensional Lie algebra, this conjecture is proved
for compact manifolds $M$ by Mishchenko and Fomenko and for non-compact manifolds
(under the assumption that all iso-energy levels $H^{-1}(h)$
are compact) by Brailov.
Then the commuting integrals can be taken as polynomials in $f_1,\dots,f_l$
(see \cite{MF3, Br2, F}).

Recently, the conjecture has been  proved in $C^\infty$--smooth case.
This means that the $r$--dimensional invariant isotropic tori $\mathbb{T}^r$
can be organized into larger, $n$--dimensional Lagrangian tori $\mathbb{T}^n$
which  are the level sets of the commutative algebra of smooth integrals
$\{g_1,\dots,g_n\}$. The point is that local commuting integrals
$\{I_1,\dots,I_r,p_1^2+q_1^2,\dots, p_k^2+q_k^2\}$ defined on different toroidal domains
can be glued smoothly on the whole manifold \cite{BJ2}.

If $\mathcal F$ is any algebra of functions
on symplectic (or Poisson) manifold, then we shall say that
$\mathcal A\subset \mathcal F$ is {\it a complete subalgebra} if
$$
\ddim{\mathcal A}+\dind{\mathcal A}=\ddim{\mathcal F}+\dind{\mathcal F}.
$$

Thus, the conjecture can be stated
as follows. Let $\mathcal F$ be a complete algebra
on a symplectic (or Poisson) manifold.
Then one can find a complete commutative
subalgebra $\mathcal A$ of $\mathcal F$, i.e.,
commutative subalgebra with differential dimension
$$
\ddim{\mathcal A}=\frac12(\ddim{\mathcal F}+\dind{\mathcal F}).
$$
Here instead of $\mathcal F$ we usually have to consider its functional extension $\tilde{\mathcal F}$,
that is, the algebra formed by the
functions $h=h(f_1,f_2,\dots,f_l)$, $f_i\in \mathcal F$, where $h$ is polynomial,
 real-analytic or $C^\infty$-smooth depending on the class of functions we want to work with.

\subsection{Integrable pairs}

We turn back to the geodesic flows on homogeneous spaces.
We have shown that the non-commutative integrability
implies the classical commutative integrability by means of
$C^\infty$--smooth integrals \cite{BJ2}.
Thus, a more delicate problem remains: the construction of complete commutative
algebras of integrals of $ds_0^2$
that are {\it polynomial in momenta}.

Let $A$ be a complete commutative algebra on generic orbits in  $\Phi(T^*(G/H))$
and let $\mathcal B$ be a complete commutative
subalgebra of $\mathcal F_2$.
Then, according to Corollary  \ref{Bol_Jov},
$\mathcal A+\mathcal B$ is a complete commutative
algebra on $T^*(G/H)$.

There is a well known construction,
called the argument shift method \cite{MF1},
which allows us to obtain a complete commutative family of polynomials on every
coadjoint orbit of a compact group.
For regular orbits this is proved by Mishchenko and Fomenko \cite{MF1}.
For singular orbits there are several different proofs:
by Mikityuk \cite{Mik1} , Brailov \cite{Br2} and Bolsinov \cite{Bo}.
(Note that $\Phi(T^*(G/H))$
can be often a subset of singular set in $\mathfrak g^*$.)
Thus, to construct a complete commutative algebra of functions on
$T^*(G/H)$ we need to find a complete
commutative subalgebra $\mathcal B$ of $G$--invariant functions
on $T^*(G/H)$.

For symmetric and weekly symmetric spaces (spherical pairs) the algebra
$\mathcal{F}_2$ is already commutative.
In a neighborhood of a generic point $x\in T^*(G/H)$ each
$G$--invariant function $f$ can be expressed as
$f=h \circ \Phi$ and
thus we can use just functions from $\mathcal F_1$ to get the integrability of
any $G$--invariant geodesic flow on $G/H$.
Spherical pairs for $G$ simple and semisimple
are  classified in \cite{Kr} and \cite{Mik3}, respectively.

To discuss the general case, one first needs to describe the
structure of the algebra $\mathcal F_2$.
Let us fix some bi-invariant metric on $G$,
i.e., $\Ad_G$--invariant scalar product $\langle\cdot,\cdot\rangle$ on $\mathfrak g$.
We can identify $\mathfrak g^*$ and $\mathfrak g$ by $\langle\cdot,\cdot\rangle$
and $T^*(G/H)$ and $T(G/H)$ by the corresponding normal metric $ds^2_0$.
Let $\mathfrak g=\mathfrak h+\mathfrak v$
be the orthogonal decomposition of $\mathfrak g$, where
$\mathfrak h$ is the Lie algebra of $H$.
Then $G$--invariant functions on $T^*(G/H)$
are in one-to-one correspondence with $\Ad_{H}$  invariant
polynomials on $\mathfrak v$.
Within this identification, the Poisson bracket on $T^*(G/H)$
corresponds to the following bracket on $\mathbb{R}[\mathfrak v]^H$
\begin{equation}
\{f(v),g(v)\}_\mathfrak v=-\langle v, [\nabla f(v),\nabla g(v) ]\rangle,
\; f,g: \mathfrak v \to {\mathbb{R}},
\label{b.0}
\end{equation}
where $\mathbb{R}[\mathfrak v]^H$ denote
the algebra of $\Ad_{H}$--invariant polynomials  on $\mathfrak v$
(see Thimm \cite{Th}).
We have
\begin{eqnarray}
&&\ddim {\mathbb{R}[\mathfrak v]^H}=
\dim \mathfrak v-\dim \mathfrak h +\dim \ann_\mathfrak h(v),\\
&&\dind {\mathbb{R}[\mathfrak v]^H}=\dim pr_\mathfrak v(\ann_\mathfrak g(v))=
\dim\ann_\mathfrak g(v)-\dim\ann_\mathfrak h(v)
\label{b.1}
\end{eqnarray}
for generic $v\in \mathfrak v$ \cite{BJ1}.
Here $\ann_\mathfrak g(v)$ and $\ann_\mathfrak h(v)$ denote
the annihilators of $v$ in $\mathfrak g$ and
$\mathfrak h$ respectively:
$\ann_\mathfrak g(v)=\{\eta\in \mathfrak g, [\eta,v]=0\}$,
$ \ann_\mathfrak h(v)=\{\eta\in \mathfrak h, [\eta,v]=0\}$.
By {\it genericity} of $v\in \mathfrak v$ we mean that the dimensions of
$\ann_\mathfrak g(v)$ and $\ann_\mathfrak h(v)$ are minimal.

From (\ref{b.1}), the condition that a commutative subalgebra
$\mathcal B\subset{\mathbb{R}[\mathfrak v]^H}$
is complete can  be rewritten in the following form:
\begin{equation}
\ddim {\mathcal B}=\frac12(\ddim{\mathbb{R}[\mathfrak v]^H}+
\dind{\mathbb{R}[\mathfrak v]^H})=\dim \mathfrak v-\frac12 \dim {\mathcal O}_G (v),
\label{b.2}
\end{equation}
for a generic $v\in \mathfrak v$, where ${\mathcal O}_G(v)$ is the adjoint
orbit of $G$.

\begin{remark} \label{dimension} {\rm
Every  Casimir function
of the algebra of $G$--invariant functions $\mathcal F_2$
in a neighborhood of a generic point $x\in T^*(\mathfrak G/\mathfrak H)$
is of the form $h\circ\Phi$,
where $h$ is a local invariant of the (co)adjoint representation.
Therefore
$$
\dind ({\mathcal F}_1 +{\mathcal F}_2)=\dind{\mathcal F}_1=\dind{\mathcal F}_2=
\ddim({\mathcal F}_1\cap{\mathcal F}_2)=\dind{\mathbb{R}[\mathfrak v]^H}.
$$
In particular,
the phase space of the geodesic flow of a normal metric is foliated by
$\dim pr_\mathfrak v(\ann_\mathfrak g(v))$--dimensional isotropic tori.
}\end{remark}

\begin{remark}\label{orbit_space}{\rm
There is a nice geometrical description of the algebra ${\mathbb{R}[\mathfrak v]^H}$.
We can pass from $\mathfrak v$ to the
orbit space $\mathfrak v/H$, with respect to the natural adjoint action
of $H$ on $\mathfrak v$.
Denote by $[v]$ the orbit of the $\Ad_H$--action through $v$.
It is clear that one can consider $\mathbb{R}[\mathfrak v]^H$
as the algebra $\mathbb{R}[\mathfrak v/H]$ of functions
on the orbit space, with respect to $\{\cdot,\cdot\}'$ -- the reduced bracket of (\ref{b.0}).
Note that $\mathfrak v/{H}$ is not smooth.
However, in a  neighborhood of a generic point,
it is a smooth manifold
of dimension $(\dim \mathfrak v-\dim \mathfrak h+\dim \ann_\mathfrak h(v))$
(minimality of dimensions of $\ann_\mathfrak h(v)$ and $\ann_\mathfrak g(v)$
means exactly that $[v]$ is a smooth point and the
bracket $\{\cdot,\cdot\}'$ has maximal rank at $[v]$).
Now, the completeness of ${\mathcal B}$ as a subalgebra
in $\mathcal F_2$ is equivalent to the condition that
${\mathcal B}$ is a complete algebra on the "singular" Poisson manifold
$(\mathfrak v/{H}, \{\cdot,\cdot\}')$.
}\end{remark}

\begin{remark} {\rm
The number $\frac12(\ddim\mathbb{R}[\mathfrak v]^H-\dind\mathbb{R}[\mathfrak v]^H)$ is equal
to the codimension $\delta(G^\mathbb{C},H^\mathbb{C})$
of maximal dimension orbits of the Borel subgroup
$B \subset G^\mathbb{C}$ in the complex algebraic
variety $G^\mathbb{C}/H^\mathbb{C}$ (see \cite{Mik4}).
The number $\delta(G^\mathbb{C},H^\mathbb{C})$
is called the {\it complexity} of
$G^\mathbb{C}/H^\mathbb{C}$ \cite{Pa}.
}\end{remark}

\begin{example} {\rm
Following \cite{MS, Mik4},
we shall say that $({G},{H})$ is an {\it almost spherical pair} if
\begin{equation}
\ddim {\mathbb{R}[\mathfrak v]^H}=2+\dind {\mathbb{R}[\mathfrak v]^H},
\label{2.1}
\end{equation}
or equivalently, if the complexity of $G^\mathbb{C}/H^\mathbb{C}$ is equal to one.
They are classified, for $G$ compact and semisimple in \cite{Pa, MS}.
As complete commutative algebra on $T^*(G/H)$
we can take arbitrary complete commutative subalgebra
$\mathcal A\subset \mathcal F_1$ and one $G$--invariant function functionally
independent of $\mathcal{F}_1$ (see Mikityuk and Stepin \cite{MS}).
The examples of Thimm \cite{Th} ($(SO(n),SO(n-2))$) and
Paternain and  Spatzier \cite{PS}
($(SU(3),\mathbb{T}^2)$) are almost spherical pairs.
In our notation, as a
complete algebra $\mathcal B\subset\mathbb{R}[\mathfrak v]^H$
we can take all the Casimir functions and an arbitrary non Casimir function
in $\mathbb{R}[\mathfrak v]^H$.}
\end{example}

\begin{definition} {\rm
We will call $(G,H)$ an integrable pair,
if there exists a complete commutative subalgebra
$\mathcal B$ in $\mathbb{R}[\mathfrak v]^H$.
}\end{definition}

We can summarize the above considerations as follows.

\begin{theorem}
If $({G},{H})$ is an integrable pair then the geodesic flow
of a normal metric
$ds^2_0$ is completely integrable in the commutative sense by means of
analytic integrals, polynomial in velocities.
\end{theorem}

Therefore, the Mishchenko--Fomenko conjecture
for the non-commutatively integrable geodesic flow
of the metric $ds^2_0$ on $G/H$ can be stated as follows.

\begin{conjecture} {\rm
All pairs  $({G},{H})$ are integrable.
}\end{conjecture}

Spherical and almost spherical pairs $(G,H)$
are simplest examples of integrable pairs.
In the section \ref{methods}, following \cite{BJ1, BJ3},
two natural methods for constructing
commutative families of $\Ad_H$ invariant functions,
namely the shift-argument method and
chain of subalgebras method, are presented. In many examples
(Stiefel manifolds, flag manifolds, orbits of the adjoint actions
of compact Lie groups etc.)
we have proved that those methods lead to complete commutative algebras.

\section{Integrable deformations of normal metrics }
\label{deformations}

Besides relation with the Mishchenko--Fomenko conjecture,
we shall explain how, with a help of commuting integrals,
one can construct new integrable geodesic flows on homogeneous spaces.

\paragraph{Submersion metrics.}
Let $A$ be a complete commutative algebra on a generic orbit in  $\Phi(T^*(G/H))$.
Then, according to Corollary \ref{Bol_Jov},
$\mathcal A+\mathcal F_2$ is a complete
algebra on $T^*(G/H)$.
Let $h_C(\xi)=\frac12\langle C(\xi),\xi\rangle$
be a quadratic positive definite polynomial in $A$.
Then $h_{C}\circ\Phi$ is the Hamiltonian of the geodesic flow of
a certain metric that we shall denote by $ds_{C}^2$.
The metric $ds^2_{C}$ has the following nice geometrical meaning.
This is the submersion metric of the right--invariant
Riemannian metric on $G$ whose Hamiltonian function is
obtained from $h_C(\xi)$ by right translations.
The geodesic flow of $ds^2_C$ is completely integrable
since $h_{C}\circ\Phi$ commutes with every function from $\mathcal A+\mathcal F_2$.
The dimension of invariant tori is equal to
$$
\dind(\mathcal A+\mathcal F_2)=\dim \mathcal O_G(v)+
\dim pr_\mathfrak v(\ann_\mathfrak g(v)),
$$
for a generic $v\in \mathfrak v$ (see \cite{BJ1}).

Notice that the argument shift method \cite{MF1} (see below) always
allows us to construct a commutative subalgebra $\mathcal A\subset\mathcal F_1$
which contains non-trivial quadratic  functions. Thus some integrable (in
non-commutative sense) deformations of $ds^2_0$ always exist.

This construction for symmetric spaces is done by Brailov \cite{Br2, Br3}.

\paragraph{G--invariant metrics.}
Let $\mathcal B$ be a complete commutative
subalgebra of $\mathcal F_2$.
Let $h_G\in\mathcal B$ be a $G$--invariant function,
quadratic in momenta and positive definite.
Then $h_G$ can be considered as the Hamiltonian of the geodesic
flow of a certain $G$--invariant metric $ds^2_G$.
The geodesic flow of $ds^2_G$ is completely integrable since
it admits the complete algebra of first integrals
$\mathcal F_1+\mathcal B$ (see Corollary  \ref{Bol_Jov}).
The dimension of invariant tori is equal to
$$
\dind(\mathcal F_1+\mathcal B)=
\dind\mathcal B=\dim \mathfrak v-\frac12 \dim {\mathcal O}_G (v),
$$
for a generic $v\in \mathfrak v$.

\paragraph{Non-invariant metrics.}
Let $h_C\circ\Phi$ and $h_G$ be as above and let
$H_{\lambda_1,\lambda_2}=\lambda_1h_C\circ\Phi+\lambda_2 h_G$
be positive definite.
Then $H_{\lambda_1,\lambda_2}$ is the Hamiltonian
of the metric which we shall denote by $ds^2_{\lambda_1,\lambda_2}$.
The family of metrics $ds^2_{\lambda_1,\lambda_2}$
has completely integrable geodesic flows, but with no obvious
symmetries. The phase space $T^*(G/H)$ is then
foliated by invariant Lagrangian tori that are level sets
of the complete commutative algebra of functions
$\mathcal A+\mathcal B$.

\begin{example}{\rm
The most natural and simplest example of integrable deformations
is as follows. Consider a compact Lie group $G$ as a homogeneous
space. The geodesic flow of the biinvariant metric $ds^2_0$ is
completely integrable in non-commutative sense and the dimension
of invariant isotropic tori in $T^*G$ is equal to ${\rm{ind}} G$.
The first integrals of the flow are all left- and right-invariant
functions on $T^*G=TG$. To obtain integrable deformations of
$ds^2_0$ one should consider a complete commutative subalgebra
$\mathcal B\subset \mathcal F_2$, where $\mathcal F_2$ is the
algebra of left-invariant functions.  Such a subalgebra can be
constructed by the argument shift. Moreover one can describe all
quadratic functions in $\mathcal B$ (see \cite{MF1}) in the
following way. Consider the linear operator $C_{a,b,D}:\mathfrak
g\to \mathfrak g$ defined by
$$
C_{a,b,D}(x)=\ad^{-1}_a \ad_b (x_1) + D(x_2),
$$
where $x=x_1+x_2$, $x_2\in \mathfrak k$, $x_1\in \mathfrak
k^\bot$, $\mathfrak k\subset\mathfrak g$ is a Cartan subalgebra,
$a, b\in\mathfrak k$ and $a$ is regular, $D:\mathfrak k\to
\mathfrak k$ is an arbitrary operator. Then we consider the
quadratic form $\frac{1}{2}\langle C_{a,b,D}(x),x\rangle$ on
$\mathfrak g$ and extend it to the whole tangent bundle $TG$ by
left translations. As a result we obtain a quadratic function
$h_{a,b,D}^{left}$ which can be considered as the Hamiltonian of a
left invariant metric on $G$. Its geodesic flow will be integrable
and the algebra of integrals consists of two parts: $\mathcal B$
(commutative part) and $\mathcal F_1$ (non-commutative part which
consists of all right-invariant functions). For such flows, the
dimension of invariant isotropic tori in $T^*=TG$ will be equal to
$\frac{1}{2}(\dim G+\ind G)$. But we can continue this deformation
procedure by choosing a complete commutative subalgebra $\mathcal
A\subset \mathcal F_1$. This can be done just in the same way as
for $\mathcal F_2$. As a result we shall obtain a right-invariant
quadratic function $h^{right}_{a',b',D'}$ which also gives an
integrable geodesic flows on $G$. But now we can take the sum
$h_{a,b,D}^{left}+h^{right}_{a',b',D'}$ which also gives an
integrable geodesic flow whose algebra of integrals $\mathcal A +
\mathcal B$ is commutative and, consequently, invariant tori are
Lagrangian, i.e., of dimension $\dim G$. }\end{example}

\section{Methods and examples}
\label{methods}

\subsection{Argument shift method}

Let $\mathbb{R}[\mathfrak g]^G$ be the algebra of
$\Ad_{G}$--invariant polynomials on $\mathfrak g$.
Mishchenko and Fomenko showed that
the polynomials
$$
{A}_a=\{p^\lambda_a=p(\cdot+\lambda a), \;\lambda\in\mathbb{R},\; p\in \mathbb{R}[\mathfrak g]^G\}
$$
obtained from the invariants by shifting the argument
are all in involution \cite{MF1}. Furthermore,
for {\it every} adjoint orbit in $\mathfrak g$,
one can find
$a\in \mathfrak g$, such that ${A}_a$ is a
complete involutive set of functions on this orbit.
For regular orbits it is proved by Mishchenko and Fomenko \cite{MF1}.
For singular orbits there are several different proofs by
Mikityuk \cite{Mik1}, Brailov \cite{Br2} and Bolsinov \cite{Bo}.

Thus, as was already mentioned, the argument shift method allows us to
construct a complete commutative subalgebra in $\mathcal F_1$.
Now we want to use it to construct such a subalgebra in $\mathcal F_2$.

By ${\mathcal B}_a$ denote the restriction of
${A}_a$ to $\mathfrak v$:
$$
{\mathcal B}_a=\{p^\lambda_a(v)=p(v+\lambda a), \;
\lambda\in\mathbb{R},\;p\in \mathbb{R}[\mathfrak g]^G,\; v\in \mathfrak v \}
$$
It can be easily checked that if $H$ is the
subgroup of the isotropy group $G_a$:
$$
H\subset G_a=\{g\in G, \; \Ad_g(a)=a\},
$$
(i.e., $\mathfrak h\subset\ann_\mathfrak g(a)$)
for some $a\in \mathfrak g$
then  ${\mathcal B}_a$ will be an algebra of
$\Ad_H$--invariant polynomials.
Also, we have the following simple observation.
If $f_1$ and $f_2$ are in involution $\{f_1,f_2\}_\mathfrak g=0$
and their restrictions to $\mathfrak v$:
$p_1=f_1\vert_\mathfrak v$, $p_2=f_2\vert_\mathfrak v$ are
$\Ad_{H}$--invariant; then $\{p_1,p_2\}_\mathfrak v=0$.
Thus, ${\mathcal B}_a$
is a commutative subalgebra of $\mathbb{R}[\mathfrak v]^H$.

In order to estimate
the number of independent functions, obtained by shifting the argument,
we look at the algebra  ${\mathcal B}_a$ from the point of view of the
bi-Hamiltonian system theory. This approach gives us a possibility to use,
in particular,
the completeness criterion proved by the first author in \cite{Bo}.

Let $\{\cdot,\cdot\}_1$ and $\{\cdot,\cdot\}_2$
be compatible Poisson structures on a manifold $N$.
Compatibility means that any linear combination of
$\{\cdot,\cdot\}_1$ and $\{\cdot,\cdot\}_2$ with constant
coefficients is again a Poisson structure. So we have a
family of Poisson structures:
$$
\Lambda=\{\lambda_1\{\cdot,\cdot\}_1+\lambda_2\{\cdot,\cdot\}_2, \;
\lambda_1,\lambda_2 \in {\mathbb{R}}, \;\lambda_1^2+\lambda_2^2\ne 0\}.
$$

By $r$ denote the rank of a generic bracket in $\Lambda$.
For each bracket $\{\cdot,\cdot\}\in\Lambda$ of rank $r$,
we consider the set of its Casimir functions.
Let ${\mathcal F}_\Lambda$ be the union of these sets.
Then ${\mathcal F}_\Lambda$ is an
involutive set with respect to every Poisson bracket from $\Lambda$.
Together with $\Lambda$, consider its natural complexification
$\Lambda^{\mathbb{C}}=\{\lambda_1\{\cdot,\cdot\}_1+\lambda_2\{\cdot,\cdot\}_2$,
$\lambda_1,\lambda_2\in \mathbb{C}, \;
\vert\lambda_1\vert^2+\vert\lambda_2\vert^2\ne 0\}$.
Here, for $\{\cdot,\cdot\}\in\Lambda^{\mathbb{C}}$,
we consider $\{\cdot,\cdot\}(x)$ as a complex valued skew-symmetric bilinear form
on the complexification of the co-tangent space $(T^*_x N)^\mathbb{C}$.

\begin{theorem} {\rm (Bolsinov, 1989 \cite{Bo})}\label{Bol}
Let $\{\cdot,\cdot\}\in \Lambda$ and $\rank \{\cdot,\cdot\}(x)=r$.
Then ${\mathcal F}_\Lambda$ is complete at x
with respect to  $\{\cdot,\cdot\}$
if and only if $\rank \{\cdot,\cdot\}'(x)=r$ for all
$\{\cdot,\cdot\}'\in \Lambda^{\mathbb{C}}$.
\end{theorem}

Note that if $\mathcal F_\Lambda$ is complete at $x$ then it is
complete in some neighborhood $U$ of $x$, or, if functions
$\mathcal F_\Lambda$ are analytic, in $N$.

Now, let us pass from $\mathfrak v$ to the
orbit space $\mathfrak v/H$ (see remark \ref{orbit_space}).
It is easy to see that the algebra ${\mathbb{R}[\mathfrak v]^H}$
is closed with respect to the $a$--bracket defined by
$$
\{f(x),g(x)\}_a=\langle a, [\nabla g(x), \nabla f(x)]\rangle.
$$
That is why $\{\cdot,\cdot\}_a$ induces a Poisson bracket
$\{\cdot,\cdot\}'_a$ on the orbit space.

Moreover, it is well known that the Lie-Poisson brackets and $a$--brackets
are compatible on $\mathfrak g$ \cite{Bo, BB}.
Thus, the brackets $\{\cdot,\cdot\}'$ and $\{\cdot,\cdot\}'_a$ are
also compatible.
Notice that $p(\cdot+\lambda a)$, where $p$ is an $\Ad_G$--invariant,
is a Casimir function of the Poisson bracket
$\{\cdot,\cdot\}'+\lambda\{\cdot,\cdot\}'_a$.
Using the above criterion (Theorem \ref{Bol}) we have found the following
conditions sufficient for ${\mathcal B}_a$ to be complete.

\begin{theorem} {\rm \cite{BJ3}}
Let $H=G_a$ for some $a\in \mathfrak g$. Suppose that there
exists generic $v\in \mathfrak v^\mathbb{C}$ such that:
\begin{eqnarray*}
(C1)&& \quad \dim \ann_{\mathfrak g^\mathbb{C}} (v+\lambda a)=
\dim\ann_{\mathfrak g^\mathbb{C}} (v),
\quad \mathrm{for\;all}\;\lambda\in\mathbb{C},\\
(C2)&& \quad
\dim \ann_{\mathfrak h^\mathbb{C}} pr_{\mathfrak h^\mathbb{C}}
([\ad^{-1}_a v,v])=\dim \ann_{\mathfrak g^\mathbb{C}} (v).
\end{eqnarray*}
Then ${\mathcal B}_a$ is a complete commutative subalgebra
in $\mathbb{R}[\mathfrak v]^H$. In particular $(G,G_a)$ is an integrable pair.
\end{theorem}

The homogeneous space
$G/G_a$ is the adjoint orbit ${\mathcal O}_G (a)$ of the $G$--action
on $\mathfrak g$.  If $a$ is regular in $\mathfrak g$ then $G_a$
is a maximal torus and $G/H$ is usually called a {\it flag manifold}.
In this simplest case conditions (C1) and (C2) can be easily verified.

\begin{corollary} {\rm (Bordemann \cite{Bor}, see also \cite{BJ1})}
Let $H$ be a maximal torus in a compact connected Lie group $G$.
If $a\in \mathfrak h$ is regular,
then ${\mathcal B}_a$ is a complete commutative subalgebra
in $\mathbb{R}[\mathfrak v]^H$. In particular $(G,H)$ is an integrable pair.
\end{corollary}

We think that conditions (C1) and (C2) hold for all compact
Lie groups and for each $a\in g$. In \cite{BJ3}
we have verified them for $U(n)$ ($SU(n)$),
and then (joint work with E. Buldaeva \cite{Bul}) for $SO(n)$ and  $Sp(n)$.

\begin{theorem}
Let G be a classical compact simple Lie group ($SO(n)$, $SU(n)$ or $Sp(n)$)
and $\mathcal O_G(a)$ be an arbitrary adjoint orbit of $G$. Then
$(G,G_a)$ is an integrable pair and ${\mathcal B}_a$ is a
complete commutative subalgebra in $\mathbb{R}[\mathfrak v]^H$.
\end{theorem}

For example, if we take $G=U(n)$ and $a\in\mathfrak u(n)$
of the form
$$
a=\diag(a_1,\dots,a_n)=\diag(\underbrace{\alpha_1,\dots,\alpha_1}_{k_1},
\underbrace{\alpha_2,\dots,\alpha_2}_{k_2},\dots,
\underbrace{\alpha_r,\dots,\alpha_r}_{k_r})
$$
then $\mathcal O_{U(n)}(a)=U(n)/U(k_1)\times\dots\times U(k_r)$ and
$(U(n),U(k_1)\times\dots\times U(k_r))$ is an integrable pair.

Pairs
$(U(n),U(k_1)\times U(k_2)\times\dots\times U(k_r))$ are integrable
for $k_1+\dots+k_r < n$ as well.
Indeed, there are commutative subalgebra $\mathfrak t\subset \mathfrak v$
and diagonal matrix $a$
such that $\mathfrak{u}(k_1)+\dots+\mathfrak{u}(k_r)+\mathfrak t=
\ann_{\mathfrak{u}(n)}(a)$.
Let $\mathbb{R}[\mathfrak t]$ be the algebra of polynomials on $\mathfrak t$
considered as functions on $\mathfrak v$.
Simple calculations show that  ${\mathcal B}_a + \mathbb{R}[\mathfrak t]$
is a complete commutative algebra of
$\Ad_{U(k_1)\times\dots\times U(k_r)}$ invariants.

Just in the same way we can prove that if the pair $(G,G_a)$ is
integrable and if $H\subset G_a$ is a normal subgroup such that $G_a/H$
is commutative, then $(G,H)$ is an integrable pair as well.

\subsection{Chains of subalgebras}

Trofimov and Thimm  devised a method for constructing functions in
involution on a Lie algebra $\mathfrak g$
by using chains of subalgebras \cite{Tr, Th}.
The idea is very simple, but important.
The invariant polynomials commute with all functions
on $\mathfrak g$. If we have some subalgebra $\mathfrak g_1\subset \mathfrak g$,
then invariant polynomials
on $\mathfrak g_1$ (naturally extended to the whole $\mathfrak g$) commute
between themselves, but also with invariant polynomials on $\mathfrak g$.
By induction, we come to the following construction.

Suppose we are given a chain of connected compact subgroups
$G_1 \subset G_2 \subset \dots \subset G_n=G$,
and the corresponding chain of subalgebras in $\mathfrak g$:
\begin{equation}
\mathfrak g_1 \subset \mathfrak g_2
\subset \dots \subset \mathfrak g_n=\mathfrak g
\label{d.1}
\end{equation}

Let $A_i$ be the algebra of
invariants on $\mathfrak g_i$ considered as a subalgebra in
$\mathbb{R}[\mathfrak g]$.  Then $A_1+\dots+A_n$ is
a commutative subalgebra of polynomials on $\mathfrak g$ \cite{Tr,Th}.

\begin{example} {\rm
The natural filtrations
$\mathfrak {so}(2)\subset \mathfrak {so}(3)\subset\dots
\subset\mathfrak{so}(n)$
and $\mathfrak {u}(1)\subset \mathfrak {u}(2)\subset\dots
\subset\mathfrak{u}(n)$
lead to complete commutative algebras on $\mathfrak{so}(n)$ and $\mathfrak{u}(n)$
respectively (see Thimm \cite{Th}).
}\end{example}

Now, if we have $H\subset G_1$ then polynomials
in $A_1+\dots+A_n$ are $\Ad_H$--invariant. Therefore
$\mathcal B=\mathcal B_1+\dots+\mathcal B_n$ is
a commutative subalgebra of $\mathbb{R}[\mathfrak v]^H$,
where $\mathcal B_i$ is restriction of $A_i$ to $\mathfrak v$.

With respect to (\ref{d.1}) we have the orthogonal decomposition
$$
\mathfrak v=\mathfrak v_1+\mathfrak v_2+\dots+\mathfrak v_n,
$$
such that $\mathfrak g_{i}=\mathfrak h+\mathfrak v_1+\dots+\mathfrak v_{i}$.
Let us define  $r_1,\dots,r_n$ by:
$$
r_i=\max_{v\in \mathfrak v_1+\dots +\mathfrak v_i}
\dim pr_{\mathfrak v_i}
\Span\{\nabla p(v),\;p\in \mathbb{R}[\mathfrak g_i]^{G_i}\}.
$$

Let  $R=r_1+\dots+r_n$.
It is clear that the number of independent functions in $\mathcal B$
is greater or equal to $R$.
Hence, if $R=\dim \mathfrak v-
\frac12\dim {\mathcal O}_G (v)$ then $\mathcal B$ is complete.
An algorithm for computing numbers $r_i$ is given by Bazaikin \cite{Ba}.

In the next theorem we give examples of some integrable pairs
with complete algebras $\mathcal B\subset \mathbb{R}[\mathfrak v]^H$
obtained by using chains of subalgebras.

\begin{theorem} \label{chains} {\rm \cite{BJ3}}
$$
\begin{array}{ll}
(SO(n),SO(k_1)\times SO(k_2)),  &\quad k_1,k_2\ge 0, k_1+k_2 \le n \\
(U(n),U(1)^{k_1}\times U(k_2)\times U(k_3)),
&\quad k_1,k_2,k_3\ge 0, k_1+k_2+k_3 \le n\\
(U(n), SO(k)), &\quad k\le n\\
(SO(n_1) \times SO(n_2), SO(k)),&\quad k \le n_1,n_2\\
(U(n_1) \times U(n_2), U(k)), &\quad k\le n_1,n_2
\end{array}
$$
are integrable pairs.
In the last two examples $SO(k)$ and $U(k)$ are diagonally embedded into
$SO(n_1) \times SO(n_2)$ and $U(n_1) \times U(n_2)$ respectively.
\end{theorem}

\begin{example}{\rm
As an example
we shall indicate the chain for Stiefel manifold $SO(m)/SO(k)$
(see \cite{BJ1}):
\begin{eqnarray*}
&&  \mathfrak g_1=\mathfrak{so}(k+1)\subset
\mathfrak g_2=\mathfrak{so}(k+2) \subset
\dots \mathfrak g_{m-k}\subset\mathfrak{so}(m), \\
&& r_i=i, \quad i=1,\dots,k, \\
&&r_{i}=\left[\frac{k+i}{2}\right],
i=k+1,k+2,\dots,m-k.
\end{eqnarray*}
}\end{example}

\begin{example}{\rm
In Theorem \ref{chains}, we consider naturally embedded subgroups
(as block matrices).
However, the same construction can be applied to some other
embeddings. As an example,
let us consider the so-called Aloff--Wallach spaces
$M_{k,l}=SU(3)/T_{k,l}$ \cite{AW}, where
$$
T_{k,l}=\left\{\left(
\begin{array}{ccc}
e^{2\pi i k\theta} & 0 & 0 \\
0 & e^{2\pi i l \theta} & 0 \\
0 & 0 & e^{-2\pi i (k+l)\theta}
\end{array}\right), \; \theta\in\mathbb{R}\right\}
k,l\in\mathbb{Z}, \vert k\vert+\vert l\vert \ne 0, kl>0.
$$

Among the spaces $M_{k,l}$ there are infinitely many with
different cohomological structures: if $k,l$ are relatively prime, then
$H^4(M_{k,l},\mathbb{Z})=\mathbb{Z}/r\mathbb{Z}$,
with $r=\vert k^2+l^2+kl\vert$ \cite{AW}.

Consider the following chain of subgroups
$G_0=T_{k,l}\subset G_1 \subset G_2=SU(3)$,
where
$$
G_1=\left\{\left(
\begin{array}{cc}
g & 0 \\
0 & \det g^{-1}
\end{array}
\right)\in SU(3), \; g\in U(2)\right\}\cong U(2).
$$

Let $\mathfrak v=\mathfrak v_1+\mathfrak v_2$
be the orthogonal decomposition as above
\begin{eqnarray*}
&&\mathfrak v_1=\left\{\left(
\begin{array}{ccc}
ia & z_{12} & 0 \\
-\bar z_{12} & ib & 0\\
0 &  0 & -i(a+b)
\end{array}
\right), \; ka+lb+(k+l)(a+b)=0\right\}, \\
&&\mathfrak v_2=\left\{\left(
\begin{array}{ccc}
0 & 0 & z_{13} \\
0 & 0 & z_{23}\\
-\bar z_{13} & -\bar z_{23} & 0
\end{array}
\right)\right\}.
\end{eqnarray*}

The algebra of functions $\mathcal{B}=\mathcal{B}_1+\mathcal{B}_2$
is complete.
Indeed, $\mathcal B_1$ is generated by
$f_1=a+b$ and $f_2=\langle v_1,v_1 \rangle$ ($r_1=2$),
$\mathcal B_2$ is generated by
$f_3=\langle v_1+v_2,v_1+v_2 \rangle$ and
$f_4=tr(v_1+v_2)^3$ ($r_2=2$)
and $2+2=4=\dim \mathfrak v-\frac12\mathcal{O}_{SU(3)}(v)=7-3$.

Aloff and Wallach proved that the $SU(3)$--invariant Riemannian metrics $ds^2_t$
on $M_{k,l}$
obtained from the quadratic forms
$$
B_t(v,v)=(1+t)\langle v_1,v_1 \rangle + \langle v_2,v_2 \rangle,
\quad v=v_1+v_2, \; v_i\in \mathfrak v_i, \quad -1< t < 0
$$
have positive sectional curvature.
Since the Hamiltonian functions of metrics $ds^2_t$
belong to $\mathcal{B}$,
the geodesic flows of $ds^2_t$ are completely integrable.
}\end{example}

\subsection{Generalized chain method}

Let $\theta$ be a Cartan involution on $\mathfrak g$ and let
$\mathfrak g=\mathfrak l+\mathfrak w$
be the corresponding orthogonal decomposition into the eigen-spaces of
$\theta$. Then $(\mathfrak g, \mathfrak l)$
is called a {\it symmetric pair} and the following relations hold:
$$
[\mathfrak l,\mathfrak l]\subset \mathfrak l,\quad
[\mathfrak l,\mathfrak w]\subset \mathfrak w,\quad
[\mathfrak w,\mathfrak w] \subset \mathfrak l.
$$

Consider the following algebra of polynomials on $\mathfrak g$:
\begin{equation}
{A}_{\mathfrak g,\mathfrak l}=
\{f(x)=p(\lambda l+w), \; p\in \mathbb{R}[\mathfrak g]^G,
 \;\lambda\in\mathbb{R}\},
\label{e.1}
\end{equation}
where $x=l+w$ is the orthogonal decomposition ($l\in \mathfrak l$,
$w\in \mathfrak w$).
Let $\mathbb{R}[\mathfrak l]$ be the algebra of polynomials on
$\mathfrak l$ considered as a subalgebra in $\mathbb{R}[\mathfrak g]$.

\begin{theorem} \label{gen_chains} (i) ${A}_{\mathfrak g,\mathfrak l}$ is a
commutative algebra of
polynomials in involution with polynomials from $\mathbb{R}[\mathfrak l]$,
i.e., commutative subalgebra in $\mathbb{R}[\mathfrak g]^L$.

(ii) ${A}_{\mathfrak g,\mathfrak l}+
\mathbb{R}[\mathfrak l]$
is a complete algebra
of polynomials on $\mathfrak g$.
In particular, if ${A}_\mathfrak l$
is a complete commutative subalgebra of
$\mathbb{R}[\mathfrak l]$, then ${A}_{\mathfrak g,\mathfrak l}+
{\mathcal A}_\mathfrak l$
will be a complete commutative algebra on $\mathfrak g$.
\end{theorem}

This result was first proof by Mikityuk \cite{Mik2}.
Also, Theorem \ref{gen_chains} is a special case of theorem 1.5 \cite{Bo}.
It is related to the
compatibility of the Lie-Poisson bracket $\{\cdot,\cdot\}_\mathfrak g$ and
the {\it $\theta$--bracket} defined by:
$
\{f,g\}_{\theta}(x)=\langle x, [\nabla g(x),\nabla f(x)]_\theta \rangle,
$
where $[\cdot,\cdot]_\theta$ is a new operation on $\mathfrak g$
which differs from
the standard one $[\cdot,\cdot]$ by the only property that $\mathfrak w$
is assumed
to be commutative
$
[l_1+w_1,l_2+w_2]_\theta=[l_1,l_2]+[l_1,w_2]+[w_1,l_2]
$
(for more details and related references see \cite{Bo, BB}).

Suppose we are given a chain of connected subgroups $G_1 \subset \dots \subset G_n$ and
the corresponding chain of subalgebras $\mathfrak g_1 \subset \dots \subset \mathfrak g_n.$
Furthermore, suppose that either $(\mathfrak g_i,\mathfrak g_{i-1})$
is a  symmetric pair or $\mathfrak v_i$ is a subalgebra
of $\mathfrak g_i$, for all $i=2,3,\dots,n$.
Here $\mathfrak g_i=\mathfrak g_{i-1}+v_i$, $i=2,3,\dots,n$.

Let $A_1$ be an arbitrary complete involutive set of polynomials
on $\mathfrak g_1$.
For symmetric pairs $(\mathfrak g_i,\mathfrak g_{i-1})$,
let  ${A}_i$ be the algebra
$A_{\mathfrak g_{i},\mathfrak g_{i-1}}$ considered on $\mathfrak g$.
Otherwise, if $\mathfrak v_i$ is a subalgebra,
then let ${A}_i$ be an arbitrary complete involutive
set of polynomials on $\mathfrak g_i$ lifted
to $\mathfrak g_n$
(it is clear that in this case $A_i$ commute
with $A_1+\dots+ A_{i-1}$).
By induction, using Theorem \ref{gen_chains}, we get
that $A_1+A_1+\dots +A_n$
is a complete commutative algebra of polynomials on $\mathfrak g_n$.

Now, if we want to apply the above construction
to the our problem,  we have to prove "an inductive step",
analogous to Theorem \ref{gen_chains}.
Suppose we are given an integrable pair $(L,H)$.
Let $L$ be a subgroup of $G$ such that
$(\mathfrak g,\mathfrak l)$ is a symmetric pair.
Let
$$
\mathfrak g=\mathfrak h+\mathfrak v, \quad
\mathfrak l=\mathfrak h+\mathfrak v_1, \quad
\mathfrak v=\mathfrak v_1+\mathfrak v_2
$$
be the orthogonal decompositions.
Let $\mathcal B_1$ be a complete commutative algebra of $\Ad_H$--invariants
on $\mathfrak v_1$ lifted to $\mathfrak v$ and
$\mathcal B_2$ be the restriction of algebra
$A_{\mathfrak g,\mathfrak l}$ to $\mathfrak v$:
$$
\mathcal B_2=
\{f(v_1+v_2)=p(\lambda v_1+v_2), \; p\in \mathbb{R}[\mathfrak g]^G,
 \;\lambda\in\mathbb{R}\},\quad v_1\in\mathfrak v_1,\quad
v_2\in\mathfrak v_2.
$$
From Theorem \ref{gen_chains} we get
that $\mathcal B_1+\mathcal B_2$ is a commutative
subalgebra of $\mathbb{R}[\mathfrak v]^H$.

\begin{question}
Is $\mathcal B_1+\mathcal B_2$ complete in $\mathbb{R}[\mathfrak v]^H$?
\end{question}

The following particular result holds.

\begin{theorem} \label{gen_chains2}{\rm \cite{BJ3}}
Suppose that $(L,H)$ is an integrable pair,
$G/L$ is a
maximal rank symmetric space
and  a generic $v_1\in \mathfrak v_1$ is a regular
element of $\mathfrak l$.
Then $(G,H)$ is an integrable pair and
$\mathcal B=\mathcal B_1+\mathcal B_2$ is a complete commutative
subalgebra of $\mathbb{R}[\mathfrak v]^H$.
\end{theorem}

The conditions of Theorem \ref{gen_chains2} are not necessary conditions
(for example, consider $(H,L,G)=(SO(n),SO(n+1),SO(n+2))$).
It would be interesting to prove that algebra $\mathcal B$ is
always complete.
Then, if we have a chain of subgroups as above, such that
$H$ is a subgroup of $G_1$ and that $(G_1,H)$ is an integrable pair,
we would get that $(G_n,H)$ is an integrable pair as well.
Simply, in this case we can take complete
algebra ${\mathcal B}_1+\dots+{\mathcal B}_n$, where
$\mathcal B_i$ is the restriction of $A_i$ to
$\mathfrak v$, for $i=2,3,\dots,n$, and
$\mathcal B_1$ is a complete commutative algebra for
$(G_1,H)$ considered as a subalgebra of $\mathbb{R}[\mathfrak v]^H$.

\begin{example}{\rm
We can use Theorem \ref{gen_chains2} for a maximal rank symmetric space $Sp(n)/U(n)$
and integrable pairs
$(U(n),U(k_1)\times\dots\times U(k_r))$ with $k_i\le [(n+1)/2]$,
$i=1,\dots,r$ (this
guaranties
the regularity conditions from the theorem)
to obtain the integrability of pairs
$
(Sp(n),U(k_1)\times\dots\times U(k_r)).$
}\end{example}

\begin{example}{\rm
Suppose that
$(SO(n_1), H_1)$ and $(SO(n_2),H_2)$
are integrable pairs, where $H_1=SO(k_1)\times\dots\times SO(k_{r_1})$ and
$H_2= SO(l_{1})\times\dots\times SO(l_{r_2})$.
Suppose that generic $u_1 \in \mathfrak u_1$ and $u_2\in \mathfrak u_2$ are regular
in $\mathfrak{so}(n_1)$ and $\mathfrak{so}(n_2)$
respectively. Here $\mathfrak u_i$ is the orthogonal
complement of $\mathfrak h_i$ in
$\mathfrak{so}(n_i)$, $i=1,2$.
(For example we can take integrable pairs
$(SO(2k+n),SO(k))$ and $(SO(2l+m),SO(2l))$ from Theorem \ref{chains})
Then, if $SO(n_1+n_2)/SO(n_1)\times SO(n_2)$ is a  maximal rank symmetric space
($n_1=n_2\pm 0,1$),
$
(SO(n_1+n_2), SO(k_1)\times\dots\times SO(k_{r_1})\times
SO(l_{1})\times\dots\times SO(l_{r_2}))
$
is an integrable pair.
}\end{example}

\begin{example}{\rm
Take
an integrable pair $(SO(n),SO(k_1)\times\dots\times SO(k_{r}))$ obtained by the above construction.
Then, since $SU(n)/SO(n)$ is a maximal rank symmetric space,
$(SU(n),SO(k_1)\times\dots\times SO(k_{r}))$
is an integrable pair as well.
}\end{example}

\section{Integrability and reduction} \label{int_red}

\paragraph{Submersions.}
Suppose we are given a compact Riemannian manifold $(Q,g)$ with a completely
integrable geodesic flow.
Let $G$ be a compact connected Lie group
acting freely on $Q$ by isometries.
The natural question arises:
will the geodesic flow on $Q/G$ equipped with the submersion metric
be integrable?

Let $\mathcal G_x+\mathcal H_x=T_x Q$
be the orthogonal decomposition of $T_xQ$, where $\mathcal G_x$
is the tangent space to the fiber
$G \cdot x$. By definition, the submersion metric $g_{sub}$ is given by
$$
\langle \xi_1,\xi_2 \rangle_{\rho(x)} =
\langle \bar \xi_1,\bar \xi_2 \rangle_x,
\quad \xi_i \in T_{\rho(x)}(Q/G), \; \bar \xi_i \in \mathcal H_x, \;
\xi_i=d\rho(\bar \xi_i),
$$
where $\rho: Q \to Q/G$ is the canonical projection. The vectors in
$\mathcal G_x$ and $\mathcal H_x$ are called
{\it vertical} and  {\it horizontal} respectively.

Let $\Phi: T^*Q\to \mathfrak g^*$ be the moment map of the natural
Hamiltonian $G$--action on $T^*Q$.
It is well known that the reduced symplectic
space
$(T^*Q)_0=\Phi^{-1}(0)/G$
is symplectomorphic to $T^*(Q/G)$.
If $H$ is the Hamiltonian function of the geodesic
flow on $Q$ then $H\vert_{\Phi^{-1}(0)}$
considered on the reduced space
will be the Hamiltonian of the geodesic
flow for the submersion metric.
If we identify $T^*Q$ and $TQ$ by the metric $g$, then
$\Phi^{-1}(0)$ will be
the set of all horizontal vectors $\mathcal H$.
Moreover, if $f$ and $g$ are $G$--invariant
and $\{f,g\}\vert_\mathcal H=0$ then
$f$ and $g$ descend to Poisson commuting functions
on the reduced space (see \cite{PS}).
Thus, the base space of the submersion has completely
integrable geodesic flow if enough $G$--invariant
commuting functions descend to independent functions.

Paternain and Spatzier proved that if the manifold $Q$ has
geodesic flow integrable by means of $S^1$--invariant integrals
and if $N$ is a surface of revolution, then the submersion
geodesic flow on  $Q\times_{S^1} N=(Q\times N)/S^1$ will be
completely integrable \cite{PS}. The connected sum
$\mathbb{CP}^{n}\sharp -\mathbb{CP}^{n}$ is an example (by a
different method they also constructed integrable geodesic flows
on $\mathbb{CP}^{2n+1}\sharp \mathbb{CP}^{2n+1}$).

\paragraph{Bi-quotients.}
Combining submersions and chain of subalgebras method,
Paternain and Spatzier \cite{PS} and Bazaikin \cite{Ba}
proved integrability of geodesic flows of normal metrics on
certain interesting  bi-quotients of Lie groups. The idea is as follows.

Consider a bi-quotient $G//U$ endowed with the normal metric $ds^2_0$. Let
$$
\phi: TG\to \mathfrak g\oplus \mathfrak g
$$
be the moment map of the $G\times G$
action on $G$:
$(g_1,g_2)\cdot g =g_1gg_2^{-1}$,
$(g_1,g_2)\in G\times G,\; g\in G$.
Here we identified dual spaces by the bi-invariant metric on $G$.
Let $\mathfrak u\subset \mathfrak g\oplus\mathfrak g$
be the Lie algebra of $U$ and let $\mathfrak w=\mathfrak u^{\bot}$
be its orthogonal complement.
Also, let $\mathcal U\subset \mathfrak g \cong T_e G$
be the  vertical space at the neutral element of the group.

Then the horizontal space $\mathcal H$ at the neutral element
is the orthogonal complement of $\mathcal U$ in $\mathfrak g$.
Let $\xi\in \mathcal H$ be the horizontal vector in the neutral of the group.
Then $d\phi(T_\xi\mathcal H)$ is the vector subspace of $\mathfrak g\oplus \mathfrak g$ equal to:
\begin{equation}
d\phi(T_\xi\mathcal H)=\mathfrak k_\xi=
 (\Delta \ann_\mathfrak g (\xi))^\perp \cap\mathfrak w,
\label{c.0}
\end{equation}
where $\Delta$ denotes the diagonal embedding (see \cite{PS, Ba}).

Suppose that $F=\{h_1,\dots,h_r\}$ is a commutative family of
$\Ad_U$--invariant polynomials (with respect to the Lie-Poisson
bracket) on $\mathfrak g\oplus \mathfrak g$. Then $\mathcal
F=\{\phi\circ{h_1},\dots,\phi\circ{h_r}\}$ will be a commutative
family of $U$--invariant functions on $TG$. Since we deal with
analytic functions, it is clear from (\ref{c.0}) that the number
of independent functions on the reduced space is greater or equal
to the number
\begin{equation}
\dim pr_{\mathfrak k_\xi}\Span\{\nabla h_1(\xi),\dots,\nabla h_r(\xi)\},\quad \xi\in\mathcal H.
\label{c.00}
\end{equation}

Bazaikin developed the method of computing numbers (\ref{c.00})
for the families of commuting $\Ad_U$--invariant polynomials
obtained by using chain of subalgebras \cite{Ba}.
In particular, he proved the complete integrability
of the geodesic flows of normal metrics on
the 7--dimensional  and 13--dimensional
bi-quotients of groups $G=U(3)\times U(2)\times U(1)$
$G=U(5)\times U(4)\times U(1)$
with strictly positive sectional curvature \cite{Ba}.

\paragraph{General approach.}
Now we will show how some of these results,
as well as some of the mentioned results on the integrability
of geodesic flows on homogeneous spaces, can
be obtained directly, considering the
relationships between symplectic reductions and the integrability
of Hamiltonian systems.

Let $G$ be a compact connected Lie group
with a free Hamiltonian action on a symplectic
manifold $(M,\omega)$.
Let
$\Phi: M \to \mathfrak g^*$
be the corresponding equivariant moment map.
Let $G_\eta$ be the  coadjoint action isotropy group of
$\eta\in\Phi(M)\subset \mathfrak g^*$.
By $(M_\eta,\omega_\eta)$ we denote the reduced symplectic space
$$
M_\eta=\Phi^{-1}(\eta)/G_\eta,
\quad
\omega_\eta(d\pi(\xi_1),d\pi(\xi_2))=\omega(\xi_1,\xi_2), \;
\xi_1,\xi_2\in T_x\Phi^{-1}(\eta),
$$
where $\pi: \Phi^{-1}(\eta)\to M_\eta$ is the natural projection.

Suppose we are given an integrable
$G$--invariant  Hamiltonian system $\dot x=X_H(x)$
with compact iso-energy levels $M_h=H^{-1}(h)$.
Then $M$ is foliated by invariant tori in an open dense
set that we shall denote by $\reg M$.

The following theorem is recently proved in \cite{Jo}
and in a slightly different version, independently, in \cite{Zu}.

\begin{theorem} \label{reduction}
If $\reg M$ intersects the submanifold
$\Phi^{-1}(\eta)$ in a dense set then
the reduced Hamiltonian system on $(M_\eta,\omega_\eta)$
will be completely integrable.
\end{theorem}

Note that we do not suppose that integrals of the original
system are $G$--invariant.
Also, the general construction used in the proof of the theorem
leads to smooth commuting integrals on $M_\eta$.
In particular, if the Hamiltonian system
$\dot x=X_H(x)$ is completely integrable by means of
$G$--invariant first integrals, then $G$ is a torus
and we can use original integrals to prove
the integrability of the reduced system \cite{Jo}.

\begin{remark}{\rm
It is obvious that reductions by a discrete group $\Sigma$
of symmetries have no influence on the integrability:
quasi-periodic motions on $M$ go to the quasi-periodic motions
on $M/\Sigma$. Namely, suppose that the trajectory $\gamma$ fill up densely
an $r$--dimensional torus $\mathbb{T}^r\subset M$.
Then $\pi(\gamma)\subset M/\Sigma$ will fill up densely
a torus of the same dimension: $\Sigma(\mathbb{T}^r)/\Sigma$.}
\end{remark}

Therefore we get

\begin{theorem} \label{submersion} {\rm \cite{Jo}}
Let a compact connected Lie group $G$ act freely by
isometries on a compact Riemannian manifold $(Q,g)$.
Suppose that the geodesic flow of $g$ is completely integrable.
If $\reg T^*Q$ intersects the space of horizontal vectors
$\mathcal H\cong \Phi^{-1}(0)$ in a dense set then
the geodesic flow on $Q/G$ endowed with the submersion metric $g_{sub}$
is completely integrable.
\end{theorem}

\begin{example}\label{Eschenburg} {\rm
Eschenburg constructed bi-quotients
$M^7_{k,l,p,q}=SU(3)//T_{k,l,p,q}$ endowed with the submersion
metrics $ds^2_{t,sub}$ with strictly positive sectional curvature.
Here $T_{k,l,p,q}\cong T^1 \subset T^2\times T^2$, where $T^2$ is
a maximal torus and $ds^2_t$ is a one--parameter family of
left-invariant metrics  on $SU(3)$ (see \cite{E}). One can prove
that the geodesic flows of the metrics $ds^2_t$ are completely
integrable and that we can apply Theorem \ref{submersion} to get
the integrability of the geodesic flows of the submersion metrics
on $M^7_{k,l,p,q}$. }\end{example}

\begin{remark}{\rm
It is very interesting that on all known manifolds which admit
metrics with strictly positive sectional curvatures (see Wilking
\cite{Wi}) one can find (positive sectional curvature) metrics
with completely integrable geodesic flows.}
\end{remark}

Here is a simple construction that gives examples satisfying the
hypotheses of Theorem \ref{submersion}. Suppose we are given
Hamiltonian $G$--actions on two symplectic manifolds
$(M_1,\omega_1)$ and $(M_2,\omega_2)$ with moment maps
$\Phi_{M_1}$ and $\Phi_{M_2}$. Then we have the natural diagonal
action of $G$ on the product $(M_1\times
M_2,\omega_1\oplus\omega_2)$, with moment map
\begin{equation}
\Phi_{M_1\times M_2}=\Phi_{M_1}+\Phi_{M_2}.
\label{c.1}
\end{equation}

If $(Q_1,g_1)$ and $(Q_2,g_2)$ have integrable geodesic flows, then
$(Q_1\times Q_2, g_1\oplus g_2)$ also has
integrable geodesic flow.
Using (\ref{c.1}),
one can easily see that if the $G$--actions
on $Q_1$ and $Q_2$ are almost everywhere locally free,
and free on the product $Q_1 \times Q_2$
then  a generic horizontal vector of the submersion
$Q_1 \times Q_2 \to Q_1 \times_G Q_2= (Q_1\times Q_2)/G$
belongs to $\reg T^*(Q_1 \times Q_2)$.
Thus the geodesic flow on $Q_1 \times_G Q_2$,
endowed with the submersion metric, is completely integrable.

\begin{example}{\rm
Suppose a compact Lie group $G$ acts freely by
isometries on $(Q,g)$.
Let $G_1$ be an arbitrary compact Lie group, which contains
$G$ as a subgroup. Let $ds^2_1$ be some
left-invariant Riemannian metric on $G_1$
with integrable geodesic flow.
Then $G$ acts in the natural way by isometries on $(G_1,ds^2_1)$.
Therefore,
if the geodesic flow on $Q$ is completely integrable,
then the geodesic flows on $Q\times_G Q$
and $Q \times_G G_1$ endowed with the submersion metrics
will be also completely integrable.}
\end{example}

\section{Geodesic flows on the spheres}

In this section we shall list some of the known interesting
integrable geodesic flows on the spheres.

\paragraph{Submersion metrics.}
The sphere $S^{n-1}$ and the cotangent bundle $T^*S^{n-1}$ are
given by the equations
\begin{eqnarray*}
&S^{n-1}=\{x\in\mathbb{R}^n, \; \langle q,q\rangle=1\}\\
&T^*S^{n-1}=\{(x,y)\in\mathbb{R}^{2n}\{q,p\}, \; \langle q,q\rangle=1,\langle q,p\rangle=0\}.
\end{eqnarray*}
Following section \ref{deformations} to construct integrable geodesic flows on the sphere
we can use the structure of a homogeneous space. This structure is not
unique but one can start from the simplest one $S^{n-1}=SO(n)/SO(n-1)$.
The moment map $\Phi: T^*S^{n-1}\to so(n)\cong so(n)^*$
of the natural $SO(n)$ action is then given by
$$
\Phi(q,p)=q\wedge p=qp^t-pq^t.
$$

To construct an integrable geodesic flow on $S^{n-1}$ we can use an
arbitrary integrable system on the Lie algebra $so(n)$ with quadratic
positive definite hamiltonian (speaking more precisely such a system must
be integrable on singular orbits lying in the image of the momentum
mapping). There are several series and some exceptional
examples of such systems (see, for instance, \cite{Ma, AM, AM2, MF1})
The most famous among them is the Manakov integrable case \cite{Ma}.  Using it, Brailov
obtained the following integrable geodesic flow \cite{Br2, Br3}.

Let $a_1>a_2>\dots>a_n>a_{n+1}$, $b_{n+1}>b_1>b_2>\dots>b_n$.

\begin{theorem} {\rm (Brailov, 1983)}
The geodesic flow of the metric on $S^{n-1}$
obtained by submersion from the right--invariant Manakov metric on
$SO(n)$ with Hamiltonian function
$$
H_{a,b}(q,p)=\frac12\sum_{1\le i<\le j\le n}\frac{b_i-b_j}{a_i-a_j}(q_ip_j-q_jp_i)^2
$$
is completely integrable. Moreover, the deformation of the metric
given by the Hamiltonian function
$F(q,p)=H_{a,b}+\frac12\sum_{1\le i\le n}\frac{b_{n-1}-b_i}{a_i-a_{n+1}}p_i^2$
is also completely integrable.
The later has integrals
$$
F_k=\sum_{1\le i<j\le n}\frac{a^k_i-a_j^k}{a_i-a_j}(q_ip_j-q_jp_i)^2-
\sum_{1\le i\le n}\frac{a_{n+1}^k-a_i^k}{a_{n+1}-a_i}p_i^2.
$$
\end{theorem}

It can be checked that the metric which corresponds to  $H_{a,b}$ with $b_i=-a_i^{-1}$
is
\begin{equation}
ds^2_a=\frac{1}{\langle A^{-1}q,q\rangle}\langle Adx,dx \rangle,
\quad
A=\mathrm{diag}(a_1,\dots,a_n).
\label{sp1}
\end{equation}

For $n=3$, the metric (\ref{sp1}) is proportional to
the metric on the Poisson sphere $S^{2}$, i.e., to the metric
obtained after $SO(2)$ reduction of the free rigid body motion
around a fixed point with inertia tensor $I=A^{-1}$ (e.g., see \cite{Bg})

Let us note that in the above construction one can use the other
representations of $S^{n-1}$ as a homogeneous space, for example,
$S^{4n-1}=Sp(2n)/Sp(2n-1)$ and $S^6=G_2/SU(3)$. It is an interesting question
whether one really can construct in such a way new examples of
integrable geodesic flows or the flows so obtained will be reduced in some
sense to the above case $SO(n)/SO(n-1)$?

\paragraph{The Maupertuis principle and Neumann system.}
Consider the natural mechanical system with Hamiltonian
$H(p,x)=\frac12\sum g^{ij}(x)p_ip_j+V(x)$ on a compact Riemannian
manifold $(M,ds^2)$. Here $g^{ij}$ is the inverse of the metric
tensor and $V(x)$ is a smooth potential on the configuration space
$M$. Let $h$ be greater than $\max V(x)$. By the classical {\it
Mapertuis principle} the integral trajectories of the vector field
$X_H$ coincide (up to reparametrization) with the trajectories of
another vector field $X_{H_h}$ with Hamoltonian
$H_h(x,p)=\frac12\sum\frac{g^{ij}(x)}{h-V(x)}p_ip_j$ on the fixed
iso-energy level $\mathcal E_h=\{H(x,p)=h\}=\{H_h(x,p)=1\}$ (see
\cite{Ar, BKF}). The Hamiltonian flow of $H_h$ is the geodesic
flow of the Riemannian metric
$$
ds^2_h=(h-V(x))ds^2,
$$
conformally equivalent to the original one $ds^2$.
Now, it is clear that
if we start with an integrable system such that $\mathcal E_h$
is almost everywhere foliated by invariant tori,
the geodesic flow of the metric $ds^2_h$ will be completely integrable.

This idea can be used to construct non-trivial integrable geodesic flows
on $S^n$ starting from integrable potential systems on the standard
sphere, ellipsoid or Poisson sphere. Such a potential systems have been
studied, in particular, in \cite{Bg, Wo, Dr, KBM, DJR}.

For example, consider the Neumann system on the sphere \cite{Neum}, i.e.,
the motion of a mass point on the sphere $S^{n-1}$ under the influence
of the force with potential $V(q)=\frac12\langle Aq,q\rangle$
(we take $A$ as above):
$$
H(q,p)=\frac12\langle p,p\rangle+\frac12\langle Aq,q\rangle.
$$
The algebraic form of the integrals
is found by K. Uhlenbeck \cite{Mo1, Mo2}:
$$
F_k(q,p)=q_k^2+\sum_{i\ne k}\frac{(q_k p_i-q_i p_k)^2}{a_k-a_i}.
$$
Therefore, the geodesic flow of the metric
$ds^2_h=(h-\langle Aq,q\rangle)\langle dq,dq\rangle\vert_{S^{n-1}}$
is completely integrable. The integrals are given by
(see \cite{BKF})
$$
\bar{F}_k(q,p)=\frac{\langle p,p\rangle}{2h-\langle Aq,q\rangle} q_k^2+
\sum_{i\ne k}\frac{(q_k p_i-q_i p_k)^2}{a_k-a_i}.
$$

Note that there is a remarkable correspondence between the  Neumann system
and the geodesic flow on the ellipsoid via the Gauss mapping
(Kn\"orrer \cite{Knorr1}).

\paragraph{Geodesical equivalence.}
After changing the coordinates $x_i=\sqrt{a_i}q_i$, the metric (\ref{sp1})
take the form
\begin{equation}
ds^2_a=\frac{1}{\langle A^{-1}x,A^{-1}x\rangle}\langle dx,dx \rangle\vert_Q,
\label{sp2}
\end{equation}
conformally equivalent to the standard metric $ds^2=\langle
dx,dx\rangle$ of the ellipsoid $Q=\{\langle A^{-1}x,x\rangle=1\}.$
There is an interesting relation between these metrics from the
point of view of geodesic equivalence. Namely, the standard metric
is geodesically equivalent to the metric (see \cite{MT, To}):
$$
d{\bar s}^2=\frac{1}{\langle A^{-1}x,A^{-1}x\rangle}\langle A^{-1} dx,dx \rangle\vert_Q.
$$
Let $g$ and $\bar g$ be the corresponding metric tensors.
Then one can define the operator
$S=\left(\frac{\det\bar g}{\det g}\right)^{\frac{1}{n}}\bar g^{-1} g$
and metrics $g_k=g S^k$, ${\bar g}_k=\bar g S^k$, $k\in\mathbb{Z}$.
It appears that metrics $g_k$ and ${\bar g}_k$ are geodesically equivalent \cite{To}.
They are all separable in elliptic coordinates
and have integrable geodesic flows. Explicit calculations
shows \cite{To}
$$
S=\left(A-x\otimes x\right)\vert_Q
$$
and, therefore, the metric ${\bar g}_1$ is given by (\ref{sp2}).
It is interesting  that the Beltrami-Klein metric of the Lobachevsky space
$\mathbb{H}^{n-1}=\{x\in \mathbb{R}^{n-1}, \; \frac{x_1^2}{a_1}+\dots+\frac{x_n^2}{a_{n-1}}\le 1\}$
can be seen as a limit of the metric $d\bar{s}^2$ as the smallest semiaxis $a_{n}$
of the ellipsoid $Q$ tends to zero \cite{DJR}.

\paragraph{Kovalevskaya and Goryachev--Chaplygin metrics on the sphere $S^2$.}
We already know that on the sphere $S^2$ we can find metrics with
integrable geodesic flows by means of an integral polynomial in
momenta of the first or second degree.
The natural question is the
existence of metrics with polynomial integral which can not be
reduced to linear and quadratic ones.
The positive answer for additional integrals of 3-th and 4-th
degrees is given by Bolsinov and Fomenko with two examples:
the Kovalevskaya $ds^2_K$ and Goryachev--Chaplygin $ds^2_{GC}$ metrics
(see \cite{BKF}).

The motion of a rigid body about a fixed point in the presence of the gravitation
field admits $SO(2)$--reduction (rotations about the direction of gravitational field).
Taking the integrable Kovalevskaya and Goryachev--Chaplygin cases we get the
integrable systems on $T^*S^2$.
The metrics $ds^2_K$ and $ds^2_{GC}$
then can easily be constructed by means of the Maupertuis principle.
They are the restrictions of the metrics
\begin{eqnarray*}
&&ds^2_{K}=\frac{h-q_1}{2}\frac{\langle Adx,dx \rangle}{\langle A^{-1}q,q\rangle},
\quad A=\mathrm{diag}(1,1,2),\\
&&ds^2_{GC}=\frac{h-q_1}{4}\frac{\langle Adx,dx \rangle}{\langle A^{-1}q,q\rangle}
\quad A=\mathrm{diag}(1,1,4)
\end{eqnarray*}
to the unit sphere.
The geodesic flow of $ds^2_K$ and $ds^2_{GC}$ admits
the first integrals of degree four and three in velocities,
which can not be reduced to lower degrees
(for more details see \cite{BKF}).

New families of metrics with cubic and
fourth degrees integrals are given by Selivanova \cite{S1,S2}.
Just recently K.Kiyohara has constructed integrable geodesic flows with
polynomial integrable of arbitrary degree $k>2$ \cite{Ki2}.
The idea of his
construction is the following. First take the constant curvature metric
$ds^2_0$. Its geodesic flow admit three independent linear integrals. Take
two of them $f_1$ and $f_2$ and consider the polynomial integral $P=f_1^l
f_2^m$  of degree $k=m+l$. It appears one can perturb the metric $ds^2_0$
in such a way that its geodesic flows remains integrable and, moreover,
the first integral $\tilde P$ preserve its form, that is, $\tilde
P=\tilde f_1^l \tilde f_2^m$ where $\tilde f_1$ and $\tilde f_2$ are
linear functions (but not integrals anymore). However, the geodesic flow
still has the property that all geodesics are closed with the same period.

\subsection*{Acknowledgments}
We are grateful to the referee for various very useful suggestions which
improved the exposition of the paper.
Also, we would like to use this opportunity to
thank the organizers of the workshop Contemporary Geometry and Related
Topics for their hospitality.
The first author was supported by Russian Found for Basic Research
(grants 02-01-00998 and 00-15-99272). The second author
was supported by the Serbian Ministry of Science and
Technology, Project 1643 (Geometry and Topology of Manifolds and Integrable Dynamical Systems).

\end{document}